\documentclass[11pt,draftclsnofoot,peerreviewca,letterpaper,onecolumn]{IEEEtran}

\usepackage{cite}
\usepackage{graphicx,color,epsfig,rotating}
\usepackage{amsfonts,amsmath,amssymb,bbm}
\usepackage{algorithm, algorithmic}
\usepackage{subfigure}
\usepackage{amsmath}
\usepackage{cite}
\usepackage{mdwtab}
\usepackage{subfigure}
\usepackage{placeins}
\usepackage{psfrag, graphicx}
\usepackage[latin1]{inputenc}
\usepackage{amssymb}
\usepackage{makeidx}

\long\def\comment#1{}

\newfont{\bbb}{msbm10 scaled 700}

\newfont{\bb}{msbm10 scaled 1100}

\newcommand{\PP}{\mbox{\bb P}}

\newcommand{\EE}{\mbox{\bb E}}

\newcommand{\Fc}{{\cal F}}

\newcommand{\Tc}{{\cal T}}
\newcommand{\Uc}{{\cal U}}

\newcommand{\SNR}{{\sf SNR}}

\newcommand{\be}{\begin{equation}}
\newcommand{\ee}{\end{equation}}
\newcommand{\bea}{\begin{eqnarray}}
\newcommand{\eea}{\end{eqnarray}}

\newtheorem{defn}{Definition}
\newtheorem{theorem}{Theorem}

\begin{document}

\title{Wireless Device-to-Device Caching Networks: Basic Principles and System Performance}

\author{Mingyue Ji,~\IEEEmembership{Student Member,~IEEE}, 
Giuseppe Caire,~\IEEEmembership{Fellow,~IEEE}, \\
and Andreas F. Molisch,~\IEEEmembership{Fellow,~IEEE}
\thanks{The authors are with the Department of Electrical Engineering,
University of Southern California, Los Angeles, CA 90089, USA. (e-mail: \{mingyuej, caire, molisch\}@usc.edu)}
}

\maketitle

\thispagestyle{empty}
\pagestyle{empty}

\vspace{-0.5cm}

\begin{abstract}
As wireless video is the fastest-growing form of data traffic, 
methods for spectrally efficient on-demand wireless video streaming are essential to both service providers and users. 
A key property of video on-demand is the {\em asynchronous content reuse}, such that a few popular files account for a large part of the traffic, 
but are viewed by users at different times.  
Caching of content on wireless devices in conjunction with device-to-device (D2D) communications allows to exploit 
this property, and provide a network throughput that is  significantly in excess of both the conventional approach of unicasting 
from cellular base stations and the traditional D2D networks for ``regular" data traffic. 
This paper presents in a {\em tutorial and concise form} some recent results on the throughput scaling laws 
of wireless networks with caching and asynchronous content reuse, contrasting the D2D approach with a few other competing approaches including
conventional unicasting, {\em harmonic broadcasting} and a novel {\em coded multicasting} approach based on 
combinatorial cache design and network coded transmission from the cellular base station only. 
Somehow surprisingly, the D2D scheme with spatial reuse and simple decentralized random caching achieves the same 
near-optimal throughput scaling law as coded multicasting. Both schemes achieve an unbounded throughput gain (in terms of scaling law) with respect to
conventional unicasting and harmonic broadcasting, in the relevant regime  where the number of video files in the library is smaller than the total size of the 
distributed cache capacity in the network. 
In order to better understand the relative merits of these competing approaches, 
we consider a holistic D2D system design that incorporates traditional microwave (2 GHz) as well as millimeter-wave (mm-wave) D2D links;  the direct connections to the base 
station can be used to provide those rare video requests that cannot be found in local caches.  We provide extensive simulation results under a variety 
of system settings, and compare our scheme with conventional unicasting, harmonic broadcasting and 
coded multicasting. We show that, also in realistic conditions and non-asymptotic regimes,  
in realistic conditions and non-asymptotic regimes the proposed  D2D approach offers very significant throughput 
gains with respect to the base station-only schemes. 
\end{abstract}



\section{Introduction}
\label{section: intro}

Wireless data traffic has been dramatically increasing over the past few years. 
Mainly driven by on-demand video streaming, it is expected to further grow from today's level by almost two orders of magnitude 
in the next five years \cite{cisco66}.  Traditional approaches for coping with this growth of demand are increasing spectral resources (bandwidth), 
spectral efficiency (modulation, coding, MIMO), or spatial reuse (density of base stations). 
However, these methods either provide only limited throughput gains in practical conditions \cite{annapureddy2010coordinated, irmer2011coordinated} 
or are expensive to implement. In particular, while heterogeneous networks with a large number of small cells can provide high 
area spectral efficiency \cite{6476878}, the necessity for high-speed backhaul connecting such large number of 
small cell base stations makes this option prohibitively expensive. 

It is noteworthy that current methods for on-demand video streaming treat video like individual data sources with (possibly) adaptive rate. 
Namely, each video streaming session  is handled  as a unicast transmission, where users successively download video ``chunks''\footnote{Typically a video chunk corresponds to 
0.5s to 1s of encoded video, i.e., to a group of pictures (GOP) between 15 and 30 frames for a typical video playback rate of 30 frames per second.} 
as if they were web-pages,  using HTTP, with possible adaptation the video quality according to the conditions of the underlying TCP/IP connection 
(e.g., Microsoft Smooth Streaming and Apple HTTP Live Streaming \cite{sanchez2011improved,sanchez2011idash,begen2011watching}) 
This approach does not exploit one of the most important properties of video, namely, a constrained request pattern, in other words, the same video is requested 
by different users, though the requests usually occur at different times.  For example, video services such as Amazon or Netflix 
provide a finite (albeit large) library of video files to the users,  and in some cases, may shape the request pattern by making some videos available 
free of charge.  It should also be noted that {\em naive multicasting} by overhearing, i.e., by exploiting the broadcast nature of the wireless medium, 
is basically useless for wireless video on-demand. In fact, while the users' demands exhibit a very significant {\em content reuse} (i.e., the same popular files are 
requested over and over), the asynchronism between such demands is so large that the probability that two users are streaming the same file 
at the ``same time'' (i.e., within a relative delay of a few seconds) is basically zero. We refer to this very peculiar feature of video on-demand as 
the {\em asynchronous content reuse}. 

Over the years, a number of other suggestions have been made to make better use of constrained 
request patterns:  \cite{li2012three, jakubczak2010softcast, aditya2011flexcast, bursalioglu2011lossy} 
considers the case  that users want the same video at the same time (e.g., in a live streaming service) 
but with a different channel quality or requested video quality. In this case, scalable video coding can be coupled with some form of broadcast 
channel coding \cite{cover2006elements}. Specifically,  scalable rateless codes \cite{shokrollahi2006raptor, luby2006raptor, oyman2010toward} 
to support heterogeneous users in a broadcast channel 
scenario are  considered in \cite{li2012three, aditya2011flexcast, bursalioglu2011lossy}. Another set of recent works considers the case where
neighboring wireless users want the {\em same} video at the same time, and collaborate in order to improve their aggregate downlink throughput.
In particular, \cite{keller2012microcast} suggests that different users download simultaneously different parts of the same video file 
from the serving base station and then share them by using device-to-device (D2D) communications. 

The above approaches are suited for synchronous streaming of live events (e.g., live sport events) but yield no gain in the presence of asynchronous content reuse, characteristic of on-demand video streaming. On the other hand, treating each user request as independent data yields a fundamental bottleneck: in conventional unicasting from a single serving base station, 
the per-user throughput decreases linearly with the number of  users in the system.
In \cite{juhn1997harmonic, paris1998efficient, engebretsen2006harmonic, paris1998low}, a coding scheme referred to as 
{\em harmonic broadcasting} is introduced.  This scheme can handle asynchronous users requesting the {\em same} 
video at different times, such that each user can start playback within a small  delay from its request time. 
With harmonic broadcasting, a video encoded at rate $R$ requires a total downlink throughput of $R \log (L/\tau)$, where $L$ is the total length of the video file and $\tau$ is the maximum playback delay.  For $\tau \ll L$ (as it is required in on-demand video streaming), 
the bandwidth expansion incurred by harmonic broadcasting can be very significant. 
%


Recent work by the authors, as well as by other research groups, has shown that one of the most promising approaches 
relies on \emph{caching}, i.e.,  storing the video files in the users' local caches and/or in dedicated helper nodes distributed in the network 
coverage area.  From the results of \cite{ DBLP:journals/corr/abs-1109-4179, ji2013optimal, gitzenis2012asymptotic}, we observe that caching can 
give significant (order) gains in terms of throughput. Intuitively, caching provides a way to exploit the inherent content reuse of on-demand video streaming 
while coping with asynchronism of the demands. Also, caching is appealing since it leverages the wireless devices storage capacity, which is 
probably the cheapest and most rapidly growing  network resource that, so far, has been left almost untapped. 

One possible approach consists of ``Femtocaching'', i.e., of deploying a large number of dedicated ``helper nodes", which cache popular video files and serve 
the users' demands through local short-range links. Essentially, such helper nodes are small base stations that use caching in order to replace the backhaul, and thus obviate 
the need for the most expensive part of a small cell infrastructure \cite{DBLP:journals/corr/abs-1109-4179}. 
Another recently suggested method combines caching of files on the user devices with a common multicast transmission of 
network-coded data \cite{maddah2012fundamental}.  We refer to this approach as {\em coded multicasting}.  The third approach, which is at the center 
of this paper, combines caching of files on the user devices with short-range device-to-device (D2D) communications \cite{ji2013optimal}. 
In this way, the caches of multiple devices form a {\em common virtual cache} that can store a large number of video files, 
even if the cache on each separate device is not necessarily very large. Both coded multicasting and D2D caching have a common interesting feature: the 
common virtual cache capacity grows linearly with the number of users in the system. This means that, as the number of users in the network grows, also
their aggregate cache capacity grows accordingly. We shall see that, qualitatively, this is the key property that allows for significant gains with respect to
the other methods reviewed here, where the content is only stored in the network infrastructure (in the serving cellular base station for conventional unicasting and harmonic
broadcasting, or in the helper nodes in  Femtocaching). 


The purpose of this paper is two-fold. On one hand, we provide a tutorial overview of the schemes and recent results on wireless on-demand video streaming 
summarized above, in terms of their throughput vs. outage probability tradeoff, in the regime where both the number of users in the system and the 
size of the library of video files grow large. 
While the results presented in Section \ref{sec: Literature Review} are not new, they have been established mostly in individual papers with different assumptions and notations; the tutorial summary presented in Section \ref{sec: Literature Review} is intended to allow a fast and fair comparison under idealized settings. 
On the other hand, looking at throughput-outage tradeoff scaling laws for idealized network models does not tell
the whole story about the relative ranking of the various schemes. Hence, in this work we present a detailed and realistic model of a single cell system with $n$ users, 
each of which has a cache memory of $M$ files, and place independent streaming requests to a library of $m$ files. 
Requests can be served by the cellular base station, and/or by D2D links. We make realistic assumptions on the channel models for the cellular links and
the D2D links, assuming that the former uses a 4th generation cellular standard \cite{sesia-LTE} and the latter 
use either microwave or mm-wave communications depending on availability \cite{azar201328, daniels201060}. 
%
By means of extensive simulations, this paper relaxes some restrictive assumptions of the theoretical scaling laws analysis based on the 
``protocol model'' of \cite{gupta2000capacity},  and provides more in-depth {\em practical} results with the goal of assessing the true potential of the various methods
in a realistic propagation environment, where the actual transmission rate of each link depends on physical quantities such as
pathloss, shadowing, transmit power and interference. 
Furthermore, we study how the use of short-range mm-wave links can influence the overall capacity. 
Such links can provide very high rates but suffer from high outage probability in some environments such as office environment (see Section \ref{sec: Simulations}).   
We investigate a composite scheme that combines robust microwave D2D links with high-capacity mm-wave links in order to achieve, opportunistically, 
excellent system performance. We also show that the type of environment in which we operate, while irrelevant for the asymptotic 
scaling laws analysis, plays a major role for the actual system throughput and outage probability. Eventually, we shall show that,
in such realistic conditions, the D2D caching scheme largely outperforms all other competing schemes
both in terms of per-user throughput and in terms of outage probability. 

The paper is organized as follows. Section~\ref{sec: Literature Review} presents a literature review 
of the recent results on wireless caching networks, where the system model and the main theoretical results are summarized.  
Then the system design approach is presented in Section~\ref{sec: System Design} and the simulation 
results are given in Section~\ref{sec: Simulations}. Conclusions are pointed out in Section~\ref{sec: Conclusions}.

\section{Literature Review}
\label{sec: Literature Review}

In this section we review the most important recent results on the throughput of wireless caching networks. 
The emphasis lies on results that use caching in combination with D2D communications, though we also review results for caching combined 
with BS-only transmission, as well as pure D2D communication (without caching). 

\subsection{Conventional Scaling Laws Results of Ad Hoc Networks and D2D communications with Caching}
\label{sec: Conventional Scaling Laws}
The capacity of conventional ad hoc networks, where source-destination pairs are drawn at random with uniform probability over the network nodes, has been studied 
extensively. Under the protocol model (see Section \ref{sec: Network Model}) and a decode-and-forward (i.e., packet forwarding) 
relaying strategy, the throughput per user of such networks scales as $\Theta(\frac{1}{\sqrt{n}})$, where $n$ denotes the number of nodes (users) in the network.
While the conclusions for realistic physical models including propagation pathloss and interference 
are more variegate  \cite{gupta2000capacity, xue2006scaling, kulkarni2004deterministic, franceschetti2007closing, ozgur2007hierarchical, franceschetti2009capacity}, 
we can conclude that practical relaying schemes are limited by the same per-user  throughput scaling bottleneck of $\Theta(\frac{1}{\sqrt{n}})$ which holds for the protocol model. 
Notice that this result assumes that the traffic generated by the network is $\Theta(n)$, i.e., constant requested throughput per user. 
This does not take into account the intrinsic content reuse of video on-demand streaming. In other words, when treating each session as independent data, 
the per-user throughput vanishes as the total demanded throughput increases.  
 
Fortunately, the behavior of video-aware networks, i.e., networks designed to support video on-demand,  
can be very different from that of conventional ad hoc networks summarized above. 
For this purpose, it is useful to consider another measure of network performance called transport capacity \cite{xue2006scaling},  which is the sum over each link of the product of the throughput per link times the distance between source and destination. 
It is known that the transport capacity of ad-hoc dense networks (i.e., networks of fixed area $O(1)$ 
with node density that scales as $\Theta(n))$,  under the protocol model, or under a physical model with decode and forward relaying, scales as $\Theta(\sqrt{n})$. For random source-destination pairs, at distance $O(1)$, the throughput per link scales again as $\Theta(\frac{1}{\sqrt{n}})$ as mentioned before. On the other hand, 
if we can reduce the distance between the source (requested file) to the destination (requesting user) to the minimum distance between nodes ($\Theta(\frac{1}{\sqrt{n}})$), which corresponds to one hop, then a constant throughput  per user can be achieved.
Such a requirement means that any user has to find the requested file within its neighborhood with high probability. The reason that the throughput per user can be improved significantly is that many short distance links can co-exist by sharing the same spectrum, which can be used more and more densely as the density of the network grows. In another word, by caching the files into the network such that request can be satisfied by short-range links, the spectrum spatial reuse of the network increases linearly with the number of users. Based on this observation, it is meaningful to consider a system design based on practical one-hop D2D transmission and caching of the video files into the user devices.

\subsection{Network Model and Problem Definitions}
\label{sec: Network Model}

\begin{figure}
\centering
\subfigure[]{
\centering \includegraphics[width=7cm]{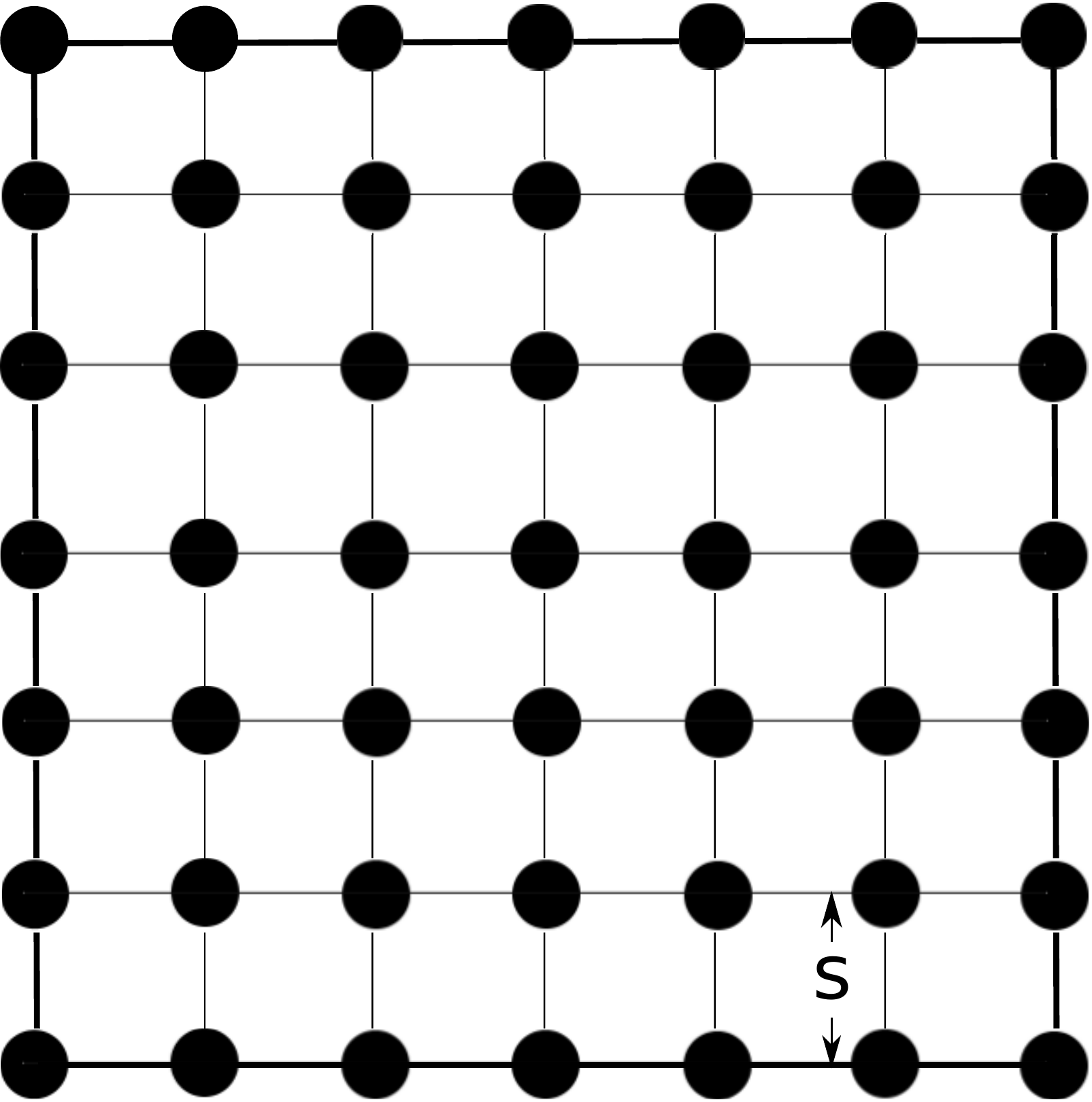}
\label{fig: Grid_Network_D2D}
}
\subfigure[]{
\centering \includegraphics[width=7cm]{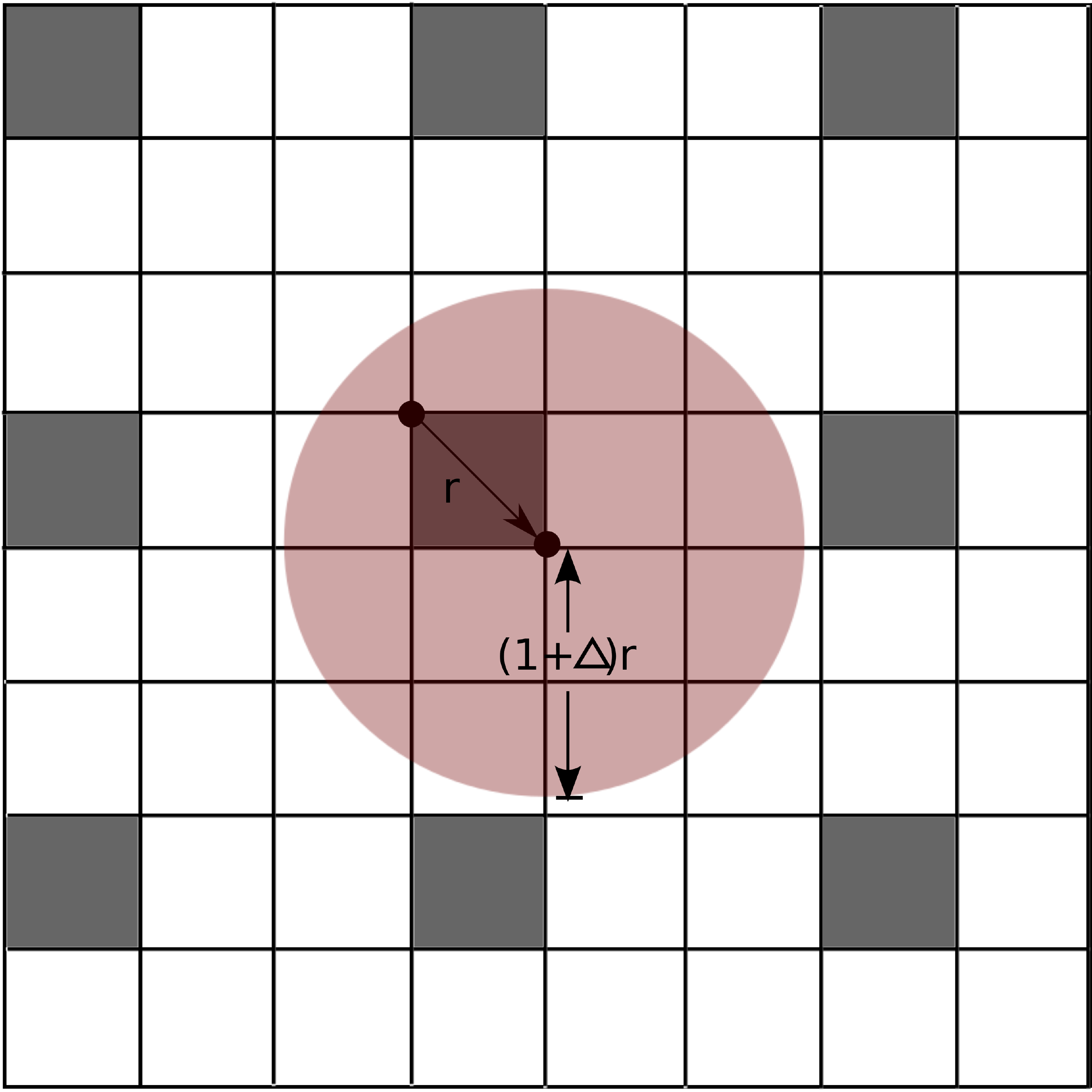}
\label{fig: Grid_TDMA}
}
\caption{a)~Grid network with $n=49$ nodes (black circles) with minimum separation $s = \frac{1}{\sqrt{n}}$. 
b)~An example of single-cell layout and the interference avoidance TDMA scheme. 
In this figure, each square represents a cluster. 
The grey squares represent the concurrent transmitting clusters. 
The red area is the disk where the protocol model allows no other concurrent transmission. 
$r$ is the worst case transmission range and $\Delta$ is the interference parameter. 
We assume a common $r$ for all the transmitter-receiver pairs. 
In this particular example, the TDMA parameter is $K=9$,  which means that each cluster can be activated every $9$ transmission scheduling slot durations.}
\end{figure}

In this section, we introduce the formal network model and the detailed problem definition for the uncoded D2D caching networks.  We consider a network formed by $n$ user nodes $\Uc = \{1, \ldots, n\}$ placed on a regular grid on the unit square, 
with minimum distance $1/\sqrt{n}$. (see Fig.~\ref{fig: Grid_Network_D2D}; for some of the later simulations, we will also consider 
the case that nodes are uniformly and randomly distributed in a $600{\rm m} \times 600{\rm m}$ square).
Each user $u \in \Uc$ makes a request for a file $f \in \Fc = \{1, \ldots, m\}$ in an i.i.d. manner, 
according to a given request probability mass function $P_r(f)$. 
Communication between user nodes obeys the protocol model \cite{gupta2000capacity}:\footnote{ In the simulations of Section \ref{sec: Simulations}, we relax the protocol model constraint and take interference into consideration by treating it like noise.} namely, 
communication between nodes $u$ and $v$ is possible if their distance $d(u,v)$ is not larger than some fixed range $r$, 
and if there is no other active transmitter within distance $(1 + \Delta) r$ from destination $v$, where 
$\Delta > 0$ is an interference control parameter. 
Successful transmissions take place at rate $C_r$ bit/s/Hz, which is a non-increasing function of the transmission range $r$ \cite{DBLP:journals/corr/abs-1109-4179}. 
In this model we do not consider power control (which would allow different transmit powers, and thus transmission ranges), for each user. 
Rather, we treat $r$ as a design parameter that can be set as a function of $m$ and $n$.\footnote{Since the number of possibly requested files $m$ typically varies with the number of users in the system $n$, and $r$ can vary with $n$, $r$ can also be a function of $m$.} 
All communications are single-hop (see also Section \ref{sec: Conventional Scaling Laws}). 
These model assumptions allow for a sharp analytical characterization of the throughput scaling law. In Section \ref{sec: Simulations}, we will see that the schemes designed by this simple model yields
promising performance also in realistic propagation and interference conditions. 

We assume that the request probability mass function $P_r(f)$ is the same for all users and follows a Zipf distribution with 
parameter $0 < \gamma_r < 1$ \cite{breslau1999web}, i.e., $P_r(f) = 
\frac{f^{-\gamma_r}}{\sum_{i=1}^m \frac{1}{i^{\gamma_r}}}$. 

We consider a simple ``decentralized'' random caching strategy, where each user caches $M$ files selected independently on the library $\Fc$ with probability $P_c(f)$. 
On the practical side, video streaming is obtained by sequentially sending ``chunks'' of video, each of which corresponds to a fixed duration.  
The transmission scheduling slot duration, i.e. the duration of the physical layer slots, is generally two to three orders of magnitude shorter than the chunk playback duration (e.g., 2 ms versus 0.5 s \cite{wu2010flashlinq}). 
Invoking a time-scale decomposition, and provided that enough buffering is used at the receiving end, 
we can always match the average throughput per user (expressed in information bit/s) with the average source coding rate at which the video file can 
be streamed to a given user.  Hence, while the chunk delivery time is fixed, 
the ``quality''  at which the video is streamed and reproduced at the user end  depends on the user average throughput.  
Therefore, in this scenario, we are concerned with the ergodic (i.e., {\em long-term average}) throughput  per user.

Referring to Fig.~\ref{fig: Grid_TDMA}, the network is divided into clusters of equal size, denoted by $g_c(m)$ (number of nodes in each cluster) 
and  independent of the users' demands and cache placement realization. A user can look for the requested file only inside its own cluster.   If a user can find the requested file inside the cluster, we say there is one \emph{potential link} in this cluster.  
We use an {\em interference avoidance} scheme for which at most one transmission is allowed in each cluster 
on any time-frequency slot (transmission resource).  
A system admission control scheme decides whether to serve potential links or ignore them.  The served potential links in the same cluster are scheduled with equal probability (or, equivalently, in round robin), such that all admitted user requests have the same average throughput $\EE[T_u] = \overline{T}_{\min}$, for all users $u$, where expectation is with respect to
the random user requests, random caching, and the link scheduling policy (which may be randomized or deterministic, as a special case).
To avoid interference between clusters, we use a time-frequency reuse scheme \cite[Ch. 17]{molisch2011wireless} with parameter $K$ 
as shown in Fig.~\ref{fig: Grid_TDMA}.  In particular, we can pick $K = \left(\left\lceil\sqrt{2}(1+\Delta)\right\rceil+1\right)^2$, where $\Delta$ is the interference parameter 
in the protocol model. 

Qualitatively (for formal definition see \cite{ji2013optimal}), we say that a user is in outage if the user cannot be served in the D2D network. This can be caused by: (i) the file requested by the user cannot be found in the user's own cluster, 
(ii) that the system admission control decides to ignore the request. We define 
the outage probability $p_o$ as the average fraction of users in outage. At this point, we can define the throughput-outage tradeoff as follows:

\begin{defn} {\bf (Throughput-Outage Tradeoff)}   \label{def: throughput-outage trade-off}
For a given network and request probability mass function $\{ P_r(f) : f \in \Fc\}$,  
an outage-throughput pair $(p,t)$ is {\em achievable} if there exists a cache placement 
scheme and an admission control and  transmission scheduling policy with outage probability
$p_o \leq p$ and minimum per-user average throughput $\overline{T}_{\min} \geq t$. 
The outage-throughput achievable region $\Tc(P_r,n,m)$ is the closure of all achievable outage-throughput pairs $(p,t)$. 
In particular, we let $T^*(p) = \sup \{ t : (p, t) \in \Tc(P_r,n,m) \}$. 
\hfill $\lozenge$
\end{defn}

Notice that $T^*(p)$ is the result of the optimization problem:
\begin{eqnarray} \label{sucaminchia}
\mbox{maximize} & & \overline{T}_{\min} \nonumber \\
\mbox{subject to} & & p_o \leq p, 
\end{eqnarray}
where the maximization is with respect to the cache placement and transmission policies. 
Hence, it is immediate to see that $T^*(p)$ is non-decreasing in $p$, since for given outage probability constraint $p_1$, $p_2$, the policies satisfying $p_2 > p_1$ are a superset of the policies satisfying $p_1$.
The range of feasible outage probability, in general, is an interval $[p_{o, \min}, 1]$ for some 
$p_{o,\min} \geq 0$.  Whether $p_{o,\min} = 0$ or strictly positive depends on the model 
assumptions.  


\subsection{Key results for D2D networks with caching}
\label{sec: Key results}

The following results are proved in \cite{ji2013optimal} and yield scaling law of the optimal throughput-outage tradeoff
under the clustering transmission scheme defined above. First, we characterize the optimal random caching distribution $P_c$:

\begin{theorem}
\label{theorem: optimal caching distribution}
Under the model assumptions and the clustering scheme, 
the optimal caching distribution $P_c^*$ that maximizes the probability that any user $u \in \Uc$ finds its requested file inside 
its own cluster is given by
\begin{equation}
\label{eq: optimal caching distribution}
P_c^*(f) = \left[1 - \frac{\nu}{z_{f}}\right]^+,  \;\;\; f = 1,\ldots, m,
\end{equation}
where $\nu = \frac{m^*-1}{\sum_{f=1}^{m^*} \frac{1}{z_{f}}}$, 
$z_{f} = P_r(f)^{\frac{1}{M(g_c(m) - 1)-1}}$, $m^* = \Theta \left (\min \{\frac{M}{\gamma_r}g_c(m), m\}\right )$ and $[\Lambda]^+ = \max[\Lambda,0]$.
\hfill  $\square$
\end{theorem}
From (\ref{eq: optimal caching distribution}), we observe a behavior similar to the water-filling algorithm for the power allocation in point-to-point communication \cite{molisch2011wireless}: if $z_{f} > \nu$, file $f$ is cached with positive probability $(1 - \frac{\nu}{z_{f}})$. Otherwise, file $f$ is not cached. 

Although the results of \cite{ji2013optimal} are more general, here we focus on the most relevant regime of the scaling of the file library size with the number of users, referred to as ``small library size'' in \cite{ji2013optimal}. Namely, we assume that $\lim_{n\rightarrow \infty} \frac{m^\alpha}{n} = 0$, where  $\alpha = \frac{1 - \gamma_r}{2 - \gamma_r}$. Since $\gamma_r \in (0,1)$, we have  $\alpha < 1/2$. This means that the library size $m$ can grow even faster than 
quadratically with the number of users $n$. In practice, however, the most interesting case is where $m$ is sublinear with respect to $n$. An example of such sublinear scaling is provided by the following simple model: suppose that user 1 has a set $m_0$ of highly demanded files, user 2 highly demanded files overlap over $m_0/2$ files with the set of user 1, and consists of 
$m_0/2$ new files, user 3 demands overlap for $2m_0/3$ over the union of user 1 and user 2, and contributes with $m_0/3$ new files and so on, such that the union of all highly demanded files of the users is $m = m_0 \sum_{i=1}^n 1/i \approx m_0 \log n$. Remarkably, any scaling of $m$ versus $n$ slower than $n^{1/\alpha}$ is captured by the following result:  

\begin{theorem} \label{theorem: 4}
In the small library regime, the achievable outage-throughput tradeoff achievable by one-hop D2D network with 
random caching and clustering transmission scheme behaves as:
\begin{align}
\label{eq: theorem 4}
&T^*(p) \geq \notag\\
& \left\{\begin{array}{ll}
\frac{C_r}{K}\frac{M}{\rho_1 m} +  \delta_1(m), & \;\; p = (1-\gamma_r) e^{\gamma_r - \rho_1} \\
\frac{C_rA}{K} \frac{M}{m (1-p)^{\frac{1}{1-\gamma_r}}} + \delta_2(m),  & \;\; p = 1 - {\gamma_r}^{\gamma_r} \left(\frac{Mg_c(m)}{m}\right)^{1-\gamma_r}, \\
\frac{C_rB}{K } m^{-\alpha} + \delta_3(m),  & \;\; 1 - {\gamma_r}^{\gamma_r} M^{1-\gamma_r} \rho_2^{1-\gamma_r} m^{-\alpha} \leq  \\
& p \leq 1 - a(\gamma_r)  m^{-\alpha},  \\
\frac{C_rD}{K} m^{-\alpha} + \delta_{4}(m),  & \;\; p \geq 1 - a(\gamma_r) m^{-\alpha}
\end{array}\right.
\end{align}
where $a(\gamma_r)$, $A, B, D$ are some constant depending on $\gamma_r$ and $M$, which can be found in \cite{ji2013optimal},
and where $\rho_1$ and $\rho_2$ are positive parameters satisfying
$\rho_1 \geq \gamma_r$ and $\rho_2 \geq \left(\frac{1-\gamma_r}{\gamma_r^{\gamma_r}M^{1-\gamma_r}}\right)^{\frac{1}{2-\gamma_r}}$. 
The cluster size $g_c(m)$ is any function of $m$ satisfying 
$g_c(m) = \omega\left(m^{\alpha} \right)$ and $g_c(m) \leq \gamma_r m/M$. The functions $\delta_i(m)$, $i = 1,2,3,4$ are vanishing for $m \rightarrow \infty$ with the following orders $\delta_1(m) = o(M/m)$, $\delta_2(m) = o\left(\frac{M}{m(1-p)^{\frac{1}{1-\gamma_r}}} \right)$,
$\delta_3(m)$, $\delta_{4}(m) = o\left(m^{-\alpha}\right)$.
\hfill $\square$
\end{theorem}
The dominant term in (3) can accurately capture the system performance even in the finite-dimensional case, as shown through simulations in \cite{ji2013optimal}. Notice that the first two regimes of (\ref{eq: theorem 4}) are the most relevant regimes in practice, providing the throughput for small outage probability. The reason of the different behaviors for these two regimes is that the first regime is achieved by a large cluster size $g_c(m)$, yielding $m^* = m$ in the optimal caching distribution. In this case, all files are stored in the common virtual cache with positive probability. 
In the second regime, $m^* < m$ if $g_c(m) < \gamma_r m/M$. The third and fourth regimes in (\ref{eq: theorem 4})  corresponds to the large outage probability regimes, where the outage probability asymptotically goes to $1$ as $m \rightarrow \infty$. These regimes may not be very interesting in practice, and are included here for completeness. 

In \cite{ji2013optimal}, we show that the throughput-outage scaling laws of Theorem 2 are indeed tight, in the sense that the analysis of an upper bound on the throughput-outage tradeoff that holds for any  one-hop scheme under the protocol model yields the  in the same order of the dominant terms with (slightly) different constants. 

\subsection{Coded Multicasting From the Base Station}
\label{sec: coded multicasting}

In this section, we review the recent work on coded multicast by the base station proposed in \cite{maddah2012fundamental}. This scheme is based on a deterministic cache placement with sub-packetization, where each user cache contains a fraction $M/m$ of packets from each of the files in the library. The scheme is designed to handle arbitrary demands. Therefore, its outage probability (under the ideal protocol model where only the base station transmits and all nodes can receive the same rate with zero packet error rate) is zero. 
Here we start with some simple example. 

We consider the case of $n = 2$ users requesting files from a library of $m = 3$ files denoted by $A$, $B$ and $C$. Suppose that the cache size of each user is $M = \frac{3}{2}$ file. Each file is divided into three packets, $A_1, A_2$, $B_1, B_2$ and $C_1, C_2$, each of size $\frac{1}{2}$ of a file. Each user $u$ caches the packets with index containing $u$. For example, user 1 caches $A_1,B_1,C_1$. Suppose that user $1$ requests $A$, user $2$ requests $B$. Then, the base station will send the packets $\{A_2 \oplus B_1\}$, where ``$\oplus$" denotes a modulo $2$ sum over the binary field), of size $\frac{1}{2}$ files, such that all requests are satisfied. Clearly, the scheme can support (with the same downlink rate) any arbitrary request. For example, suppose that user $1$ wants $B$ and user $2$ wants $C$, then the base station will send $\{B_2 \oplus C_1\}$, which again results in transmitting $\frac{1}{2}$ files.

The scheme is referred to as ``coded multicasting'' since the base station multicasts a common message to all the users, formed by linear combinations of the packets of the requested files. The term ``coded'' refer to the fact that sending linear combinations of the messages is a instance of linear network coding \cite{li2003linear, ho2006random}.

By extending this idea to general $n$, $m$ and $M$, and letting
$N_{TX}(n,m,M)$ denote the number of equivalent file transmissions form the base station, 
\cite{maddah2012fundamental}  proves the following result:
\begin{theorem}
\label{theorem: multicasting 1}
For any $m$, $n$, $M$ and arbitrary requests, for $\frac{Mn}{m} \in \mathbb{Z}^+$ and $M < m$,
\be
\label{eq: rate coded multicast}
N_{\rm TX}(n,m,M)  =  n\left(1-\frac{M}{m}\right) \frac{1}{1+\frac{Mn}{m}},
\ee
is achievable. For $\frac{Mn}{m} \notin \mathbb{Z}^+$, the convex lower envelope of the points with coordinates 
$(n,m,M,N_{\rm TX}(n,m,M))$ for integer $\frac{Mn}{m}$ is achievable.  \hfill $\square$
\end{theorem}


Through a compound channel (over the requests) and cut-set argument, [24] proves that the best possible caching and delivery scheme transmitting form the base station requires a number of transmissions not smaller than 1/12 of (4). This means that, within a bounded multiplicative gap not larger than 12, the coded multicasting scheme of [24] is information theoretically optimal, under the arbitrary demand and zero outage probability constraint, and under the protocol model. 

Letting $C_{r_0}$ 
denote the rate at which the base station (BS) can reach any point of the unit-square cell, the corresponding order-optimal per-user throughput achieved by 
the coded multicast scheme is: 
\be
\label{eq: coded multicast throughput}
T^*_{\rm BS, coded} = \frac{C_{r_0}}{N_{\rm TX}(n,m,M)}.
\ee
Obviously, since the scheme is designed to handle {\em any} user request, the outage probability of this scheme is $p_o = 0$. 

This throughput is achieved by deterministic cache placement placement. This means that when the size of library $m$ or the number of users $n$ changes, all caches must be re-arranged. To deal with the problem that $n$ is changing over time, a randomized cache placement approach was proposed in \cite{maddah2013decentralized}. For the delivery phase of this scheme, a clique covering approach is required, which is a well-known NP-complete problem. However, given the random caching structure, a greedy solution can be found and is approximately optimal within a constant factor.

\subsection{Harmonic and Conventional Broadcasting}
\label{sec: Harmonic}

\begin{figure}[ht]
\centerline{\includegraphics[width=8cm]{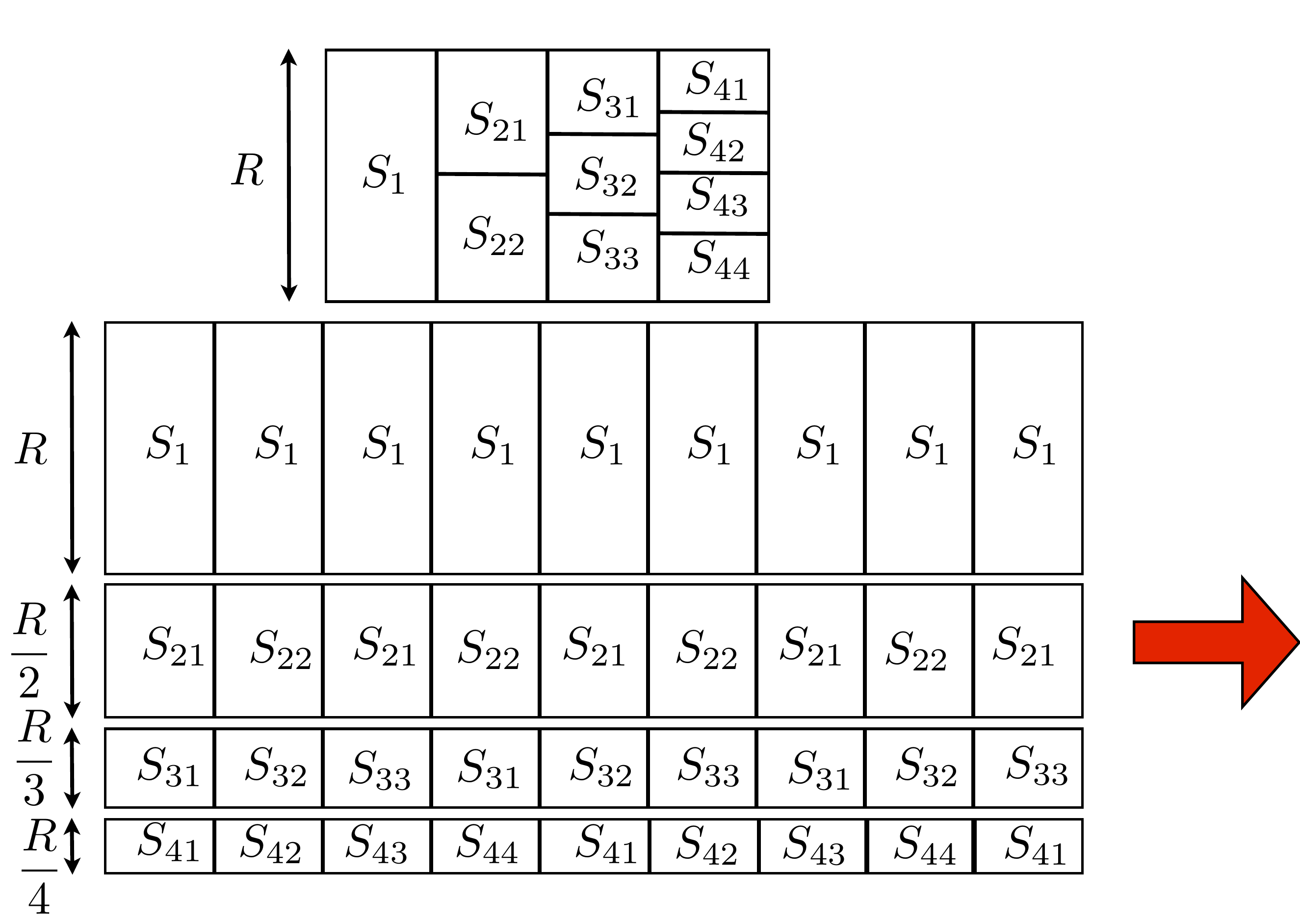}}
\caption{A video file encoded at rate $R$ is split into blocks $S_{ij} : j = 1,\ldots, i$, for $i = 1,\ldots, P$ (we have $P = 4$ in the figure), 
such that the size of $S_{ij}$ is $\tau/i$ chunks. 
Each $i$-th set of blocks is periodically transmitted in a downlink parallel channel of rate $R/i$. 
Any user tuning into the multicast transmission can start its playback after at most $\tau$ chunks.}
\label{harmonic}
\end{figure}

In brief, harmonic broadcasting works as follows: fix the maximum waiting delay of $\tau$ ``chunks'' 
(from the time a streaming session is initiated to the time playback is started), and let  $L$ denote the total length of the video file, again expressed in chunks. 
In harmonic broadcasting, the video file is split into successive blocks such that for $i = 1, \ldots, P$, there are
$i$ blocks of length $\tau/i$, where $P = \lceil L/\tau \rceil$ (see Fig.~\ref{harmonic}). Then, 
each $i$-th set of blocks of length $\tau/i$ is repeated periodically on a (logical) subchannel of throughput $R/i$, where $R$ is the transmission 
rate (in bit/s) of the video  playback (see again Fig.~\ref{harmonic}).  Users receive these channels in parallel.

In this way, each file requires a downlink 
rate of $R\log(L/\tau)$. Hence, the total number of files that can be sent in the common downlink stream is 
$m' = \min\left\{\frac{C_{r_0}}{R \log(L/\tau)}, m\right\}$.
This yields an average throughput per user of $R\left(1-p_o \right)$ with outage probability $p_o = \sum_{f = m' + 1}^m P_r(f)$ since all requests to files not included in the common downlink stream are necessarily in outage. 

Finally, the conventional approach of today's technology in cellular and wifi networks consists of handling on-demand video streaming requests exclusively at the application layer. Then, the underlying wireless network treats these requests as independent individual data. In this case, the average throughput per user is 
$\Theta\left(\frac{C_{r_0}}{n ( 1 - p_o)} (1-p_o)\right) = \Theta\left(\frac{C_{r_0}}{n}\right)$ for a system whose admission control serves a fraction 1$ - p_o$ of the users, and denies service to a fraction $p_o$ of users (outage users). 


\subsection{Summary: Comparison between Different Schemes}

In this section, we compare the schemes reviewed before in terms of theoretical scaling laws. We mainly consider four schemes, which are conventional unicasting, harmonic broadcasting, coded multicasting and the uncoded D2D caching scheme. In the following, we focus on case where $Mn \gg m$, $M$ is a constant and $m, n, L \rightarrow \infty$. As indicated in Section \ref{section: intro}, we consider a single cell of fixed area containing one base station and $n$ user nodes (dense network), and take into account adjacent cell interference into the noise floor level. 
%

For the conventional unicasting where no D2D communication is possible and the users cannot cache files, the system serves users' requests as if they were independent messages from
the BS. Hence, we are in the presence of an information-theoretic  broadcast channel with independent messages, whose per-user throughput is known to scale
as $\Theta(1/n)$, i.e., even significantly worse than the ad-hoc networks scaling law.  
As an intermediate system, we may consider the case of a conventional caching system (e.g., using prefix caching as advocated in the classical literature on caching \cite{sen1999proxy}), where users can cache $M$ files\footnote{For the sake of throughput v.s. outage, in the setting of this paper, whether the users cache $M$ most popular files, or the initial segment of a larger set of most popular files for a total size of $M$ equivalent files (prefix caching) is irrelevant.}, but the system does not handle D2D communication. 
A naive approach to this case let each user cache its $M$ most popular files.
Thus, if a user requests such a file, it will find it on its own device (this is known as the ``local caching gain"). 
However, all other files have to be downloaded from the BS in the same manner as for the conventional no-caching case.
Thus, the fundamental scaling behavior of this case is again $\Theta(1/n)$. 

In the case of harmonic broadcasting, as mentioned in Section \ref{sec: Harmonic}, if we constrain the maximum waiting time to be at most $\tau$ chunks, then the throughput per user of harmonic broadcasting scales as $\Theta\left(\frac{1}{m'\log \frac{L}{\tau}}\left(1-\sum_{f = m' + 1}^m P_r(f)\right)\right)$, where $m' \leq m$, then the throughput goes to zero as $L \rightarrow \infty$.

Next, we examine the scaling laws of the throughput for the uncoded D2D scheme for arbitrary small outage probability. By using the first line of (\ref{eq: theorem 4}), the average per-user throughput scales as $\Theta\left(\frac{M}{m}\right)$, 
which is very attractive, since the throughput increase linearly with the size of the user cache.

Finally,  from (\ref{eq: coded multicast throughput}) we observe that the throughput of coded multicasting scales as $\Theta\left(\frac{M}{m}\right)$. Interestingly, 
this scaling behavior coincides with that achieved by D2D with random caching. This indicates that by one-hop communication (either D2D or multicasting from the base station), the fundamental limit of the throughput in the regime of small outage probability is $\Theta(\frac{M}{m})$.\footnote{Notice that, in practice, also coded multicasting is subject to outages, due to the shadowing of the channel between the base station and the users. Since a common transmission rate has to be guaranteed for all the users, then some users will be in outage if the channel capacity between these users and base station is less than the common transmission rate.}

Clearly, we can see that in the regime of $Mn \gg m$, where the total network storage is larger than the library size, both the uncoded D2D caching scheme and the coded multicasting scheme have an unbounded gain with respect to the conventional unicasting and harmonic broadcasting as $m,n,L \rightarrow \infty$. 
According to the above model, uncoded D2D caching and coded multicasting are equivalent in terms of throughput scaling laws. 
However, in practice, several other factors play  a significant role in determining the system throughput and outage, in realistic conditions. For example, the availability of D2D links may depend on the specific models for propagation at short range and may significantly differ depending on the frequency band such links operate. Also, coded multicasting requires to send a common coded message to all the users in the cell. Multicasting at a common rate incurs in the worst-case user bottleneck, since in practice users have different path losses and shadowing conditions with respect to the base station. Hence, in order to appreciate the different performance of the various schemes reviewed in this paper in realistic system conditions, beyond the scaling laws of the protocol model, in the next sections we resort to a holistic system optimization and simulation in realistic conditions. 
\section{System Design}
\label{sec: System Design}




We assume that devices can operate in multiple frequency bands. For transmission from the BS to mobile stations (MS), we assume operation at 2.1 GHz carrier frequency, corresponding to one of the standard {\em long-term-evolution} (LTE) bands.\footnote{Due to the non-universal availability of sub-1GHz bands for LTE, we do not consider it further in this paper.} We furthermore assume that D2D communication can occur at 2.45 GHz carrier frequency (specifically in the Industrial, Scientific and Medical (ISM) bands), as well as in the unlicensed mm-wave band at 38 GHz. Note that the 2.45 and 38 GHz bands are not suitable for BS-to-MS communications due to propagation conditions as well as transmit power restrictions imposed by frequency regulators. Obviously, the 38 GHz band provides the possibility for very high data rates, due to the high available bandwidth at that frequency.  

\subsection{Holistic Multi-Frequency D2D System Design}
\label{sec: holistic desigen}

In this case, we try to use all the resources in the network. As discussed above, file delivery is most efficiently achieved if it involves a short range communication link, for which 
the mm-wave frequency band is ideally suited.  
However, such connections are not robust since the mm-wave can be easily blocked by walls or even human body. Hence, if the mm-wave link is not available, the next best option 
is then D2D communication in 
the $2.45$ GHz band. Finally, if even this band is not available or if the requested file is not present within the range of the D2D connections, the file may be served from the BS using the cellular downlink, depending on the admission control decisions. For the D2D links, we use clusterization, i.e., D2D communication is possible within a cluster, but not between clusters. For the cache placement, an independent and randomized cache placement of complete files is used as described in Section \ref{sec: Key results}. The clustering and caching placement algorithm is summarized in \emph{Algorithm}~\ref{algorithm: 3}\footnote{In reality, there is a chance that the same file is selected to be cached multiple times in the same cache. Although irrelevant for the sake of the throughput scaling laws, this case should be avoided in practice by practical caching algorithms. We do not further consider this aspect since its impact on the overall system performance is negligible in our simulation setting.}, in which we focus on the regime of small outage, where all potential links (requests found in the cluster) are served. 

\begin{figure}
\centering
\includegraphics[width=15cm]{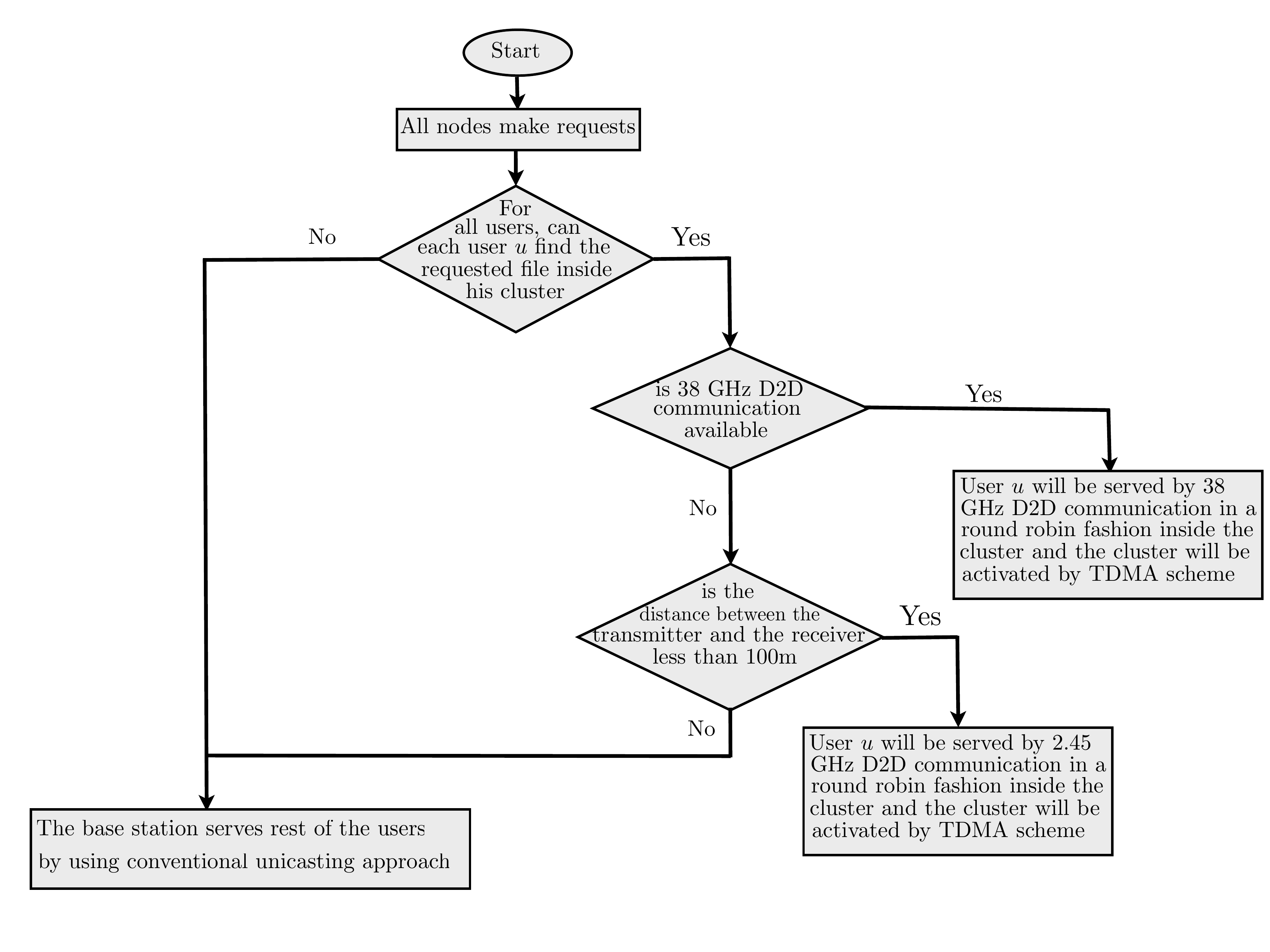} 
\caption{The flow chart of the delivery algorithm for the combination of D2D communications and multicast by the base station.} 
\label{fig: FlowChart}
\end{figure}

The flow chart of the delivery algorithm is shown in Fig.~\ref{fig: FlowChart}.
\begin{algorithm}
\caption{Clustering and uncoded caching placement}
\label{algorithm: 3}
\begin{algorithmic}[1]
\STATE According to the given outage probability of D2D communications (the probability that any user is not served by the D2D networks), decide the cluster size $g_c(m)$ by using (\ref{eq: theorem 4}). 
\FORALL {$u \in \mathcal{U}$} 
\STATE Node $u$ randomly caches $M$ files independently according the probability distribution given in (\ref{eq: optimal caching distribution}). 
\ENDFOR
\end{algorithmic}
\end{algorithm}

\subsection{Conventional Unicasitng, Coded Multicasting  and Harmonic Broadcasting Approaches}

\begin{itemize}

\item Conventional Unicasting Approach: For the cache placement, we consider that each node caches a fraction of $M/m$ of each file such that the local caching gain ($1-\frac{M}{m}$) can be obtained. 
As our baseline approach, we consider a fairness constraint subject to all the user have the same outage probability. Consider a link between BS and user $u$, with log-normal shadowing $\chi_{\sigma}$ and deterministic distance-dependent pathloss $PL(d_u)$ not including the shadowing,\footnote{In Section \ref{sec: Channel Models}, $PL$ is defined to include the log-normal shadowing $\chi_\sigma$ for the ease of presentation.} where $d_u$ denotes the distance between the BS and user $u$, let $C_u$ denote the required individual downlink rate for user $u$ such that for a fixed outage probability $p_o$, $\forall u \in \{1, \cdots, n\}$, we have
\be
\label{eq: unicasting 4}
\PP\left(\log\left(1 +  \frac{\SNR}{\chi_{\sigma}  PL(d_u)} \right) \leq C_u\right) = p_o,
\ee
where $\SNR$ (Signal-to-Noise Ratio) denotes the ratio between the transmit power and the noise power spectral density 
at the receiver. 
From (\ref{eq: unicasting 4}), the user u downlink rate $C_u$ can be determined as a function of $p_o$ and $d_u$, given the log-normal distribution of  $\chi_\sigma$. 

Now, given $C_u$, $\forall u$, let $\rho_u$ denote the fraction of downlink transmission resource dedicated to serve user $u$.
In order to maximize the minimum user rate (our reference performance metric, see (\ref{sucaminchia})), the downlink transmission resource allocation is the solution of:
\begin{eqnarray} \label{unicasting op}
\mbox{maximize} & & \min_{u \in \mathcal{A}} \rho_uC_u \nonumber \\
\mbox{subject to} & & \sum_{u \in \mathcal{A}}  \rho_u \leq 1, 
\end{eqnarray}
where $\mathcal{A}$ is the set of users that are not in outage.
It is immediate to obtain the solution of (\ref{unicasting op}) as $\rho_u = \frac{\frac{1}{C_u}}{\sum_{u' \in \mathcal{A}}\frac{1}{C_{u'}}}$. Thus, let $1_u$ be the indicator that user $u$ is not in outage, we obtain
\begin{eqnarray}
\label{eq: unicast 2}
\overline{T}_{\min} & = & \EE[\rho_u C_u 1_u] = \EE\left[\frac{\frac{1}{C_u}}{\sum_{u' \in \mathcal{A}}\frac{1}{C_{u'}}} C_u 1_u \right] \notag\\
& = & \EE\left[\frac{1_u}{\sum_{u' \in \mathcal{A}}\frac{1}{C_{u'}}}\right] = \EE\left[\frac{1_u}{\sum_{u' =1}^{n}\frac{1}{C_u'}1_{u'}}\right].  
\end{eqnarray}
Therefore, for any given $p_o$, we obtain the set of points $(\overline{T}_{\min}, p_o)$ yields the throughput-outage tradeoff achievable by conventional unicasting.  
It is immediate to see that $\overline{T}_{\min}=\Theta(1/n)$ for any target $p_o$ in $(0,1)$.


\item Coded Multicasting Approach: 
Since common multicast messages have to be decoded by all the served users, the downlink rate $R$ 
is possible if 
\be
\label{eq: coded multicasting design 1}
R = \frac{C_{r_0}}{N_{\rm TX}(n,m,M)} <  \frac{\log(1 + \frac{\SNR}{\chi_{\sigma}  PL(d_u)} )}{N_{\rm TX}(n,m,M)},
\ee
where $N_{\rm TX}(n,m,M)$ is given by (\ref{eq: rate coded multicast}). 
Since for all channel models used in our results, the receiver SNR of users at larger distance from the BS is stochastically dominated by the receiver SNR of users at smaller distance from the BS, the most stringent outage condition is imposed on the worst case users at largest distance $r_0$ from the BS, namely,  
\be
\PP \left (\frac{\SNR}{\chi_{\sigma}  PL(r_0) } > 2^{C_{r_0}}- 1 \right ) < \PP \left (\frac{\SNR}{\chi_{\sigma}  PL(d_u)} > 2^{C_{r_0}}- 1 \right ) 
\ee
for $d_u < r_0$. Hence, we have
\be
\label{eq: coded multicast rate}
\overline{T}_{\min} = R \left(1 - \PP\left( \chi_\sigma PL(r_0) \geq \frac{\SNR}{2^{C_{r_0}}-1}\right)\right)
\ee
The outage probability is given by
\begin{equation} \label{po-link} 
p_o = \frac{1}{n}\sum_{u}\PP \left (\chi_{\sigma}  PL(d_u)  \geq \frac{\SNR}{2^{C_{r_0}}- 1} \right ).
\end{equation}
Note that $C_{r_0}$ is the only control parameter, for this scheme. Hence, using (\ref{eq: coded multicast rate}) and (\ref{po-link}), the throughput-outage tradeoff of coded multicasting can be obtained by varying the common downlink rate $C_{r_0}$. 

\item Harmonic broadcasting approach: In this case, the common multicasting messages also have to be decoded by all the served users. Let $m' \leq m$, the downlink rate reliable $R$ is possible if
\be
\label{eq: harmonic design 1}
R = \frac{C_{r_0}}{m'\log \frac{L}{\tau}} <  \frac{\log(1 + \frac{\SNR}{\chi_{\sigma}  PL(d_u)} )}{m'\log \frac{L}{\tau}}.
\ee
Two events can cause outage for harmonic broadcasting: i) the physical channel is not good enough to support the video encoding rate;  ii) the requested file is not included in the set of files broadcasted by the BS. 
Since the two outage events are independent, we have
\be
\label{eq: harmonic design 2}
\overline{T}_{\min} = R \left(1 - \sum_{f=m'+1}^m P_r(f)\right) \left(1 - \PP\left(\chi_\sigma PL(r_0) \geq \frac{\SNR}{2^{C_{r_0}} - 1}\right)\right).
\ee
The outage probability is given by
\be
\label{eq: harmonic design 3}
p_o = \frac{1}{n}\sum_{u=1}^n \left(1 - \left(1 - \sum_{f=m'+1}^m P_r(f)\right) \left(1 - \PP\left(\chi_\sigma PL(d_u) \geq \frac{\SNR}{2^{C_{r_0}} - 1}\right)\right)\right).
\ee
Also in this case, by varying the common downlink rate $C_{r_0}$, by (\ref{eq: harmonic design 2}) and (\ref{eq: harmonic design 3}) we can obtain the throughput-outage tradeoff of harmonic broadcasting.

\end{itemize}

\section{Simulations and Discussions}
\label{sec: Simulations}

In this section, we provide the simulation results and some discussions. First, we describe the environments and discuss the
channel models for the three types of links mentioned in Section \ref{sec: System Design}. We then present our simulation results and discuss implications for deployment.

\subsection{Deployment Environments}

We perform simulations in two types of environments: (i) office environments and (ii) indoor hotspots. More specifically we assume a cell of dimensions  $0.36 {\rm km}^2$ ($600{\rm m} \times 600{\rm m}$) that contains buildings as well as streets/outdoor environments. We assume $n=10000$, i.e., on average, there are $2 \sim 3$ nodes, every square $10 \times 10$ meters for the grid network model. We will also investigate the effect of user density later.

For the office environment, we assume that the cell contains a Manhattan grid of square buildings with side length of $50$m, separated by streets of width $10$m. Each building is made up of offices;  of size $6.2 {\rm m} \times 6.2 {\rm m}$.\footnote{The motivation for this size stems from the line-of-sight (LOS) model (see below); we choose the office size such that it results in a LOS probability of $0.5$ if two devices are half the office dimensions apart from each other.} Corridors are not considered, as they would lead to a further complication of the channel model. 

For the ``indoor hotspot" model, which describes big factory buildings or airport halls, we also assume that the cell is filled with multiple buildings. The size of these buildings are squares with side length of $100$m and distributed on a grid with street width of $20$m. There are no internal partitions (walls) within the building. 

Within the cell, users (devices) are distributed at random according to a uniform distribution. Due to our geometrical setup, each node is assigned to be outdoors or indoors, and (in the case of the office scenario) placed in a particular office.\footnote{Note that the division of buildings into offices is only used to determine the ``wall penetration loss", while the basic pathloss and LOS probability are determined by the purely distance-dependent model (see below for details).  } This information is exploited to determine which channel model (e.g., indoor-to-indoor or outdoor-to-indoor) is applicable for a particular link. The use of such a virtual geometry is similar in spirit to, e.g., the Virtual Deployment Cell Areas (VCDA) of the COST 259 microcellular model \cite{correia2001wireless}. 

\subsection{Channel Models}
\label{sec: Channel Models}

Corresponding to the three types of transmissions (cellular, microwave D2D, millimeter-wave D2D), we use three types of channel models. We only consider pathloss and shadowing, since the effect of small scale fading can be eliminated by frequency/time diversity over the bandwidth and timescales of interest. 

The channel models are mostly obtained from the Winner II channel models \cite{winner2007d1}. We note that although these channels are not explicitly defined for device-to-device, the range of parameter validity includes device height of about $1.5$ m, which is typical for user-held device.

\subsubsection{LOS probability}

One of the key parameters for any propagation channel is the existence of a line-of-sight (LOS) condition: all channel characteristics, including path loss, delay spread, and angular spread, depend on this issue. It is obvious that the existence of a LOS is independent of the carrier frequency; it thus seems straightforward to simply apply the LOS model of Winner at all frequencies. However, as we will see in the following, there are subtleties that depend on the carrier frequency and greatly impact the overall performance. By denoting the distance between each transmitter-receiver pair as $d$, the LOS probability ($P^{W}_{{\rm LOS}}$) models by Winner \cite{winner2007d1} are summarized in Table \ref{table: LOS probability models}.

\begin{table}[ht]
\caption{The LOS probability models.}
\centering 
\begin{tabular}{c|c}
\hline\hline
indoor office  & $
P^{W}_{{\rm LOS}} =  \left\{\begin{array}{cc}1, & d \leq 2.5 \notag\\
1-0.9(1-(1.24-0.61\log_{10}(d))^3)^{\frac{1}{3}}, & d > 2.5
\end{array}\right. $ \\
\hline
indoor hotspots  &  $P^{W}_{{\rm LOS}}=\left\{\begin{array}{cc}1, & d \leq 10 \\
\exp\left(-\frac{d-10}{45}\right), & d > 10
\end{array}\right.$\\
\hline
outdoor-to-outdoor  & $P^{W}_{{\rm LOS}} = \min(18/d, 1) (1-\exp(-d/36)) + \exp(-d/36)$ \\
\hline 
outdoor-to-indoor  & $P^{W}_{{\rm LOS}} = 0$ \\
\hline 
\end{tabular}
\label{table: LOS probability models}
\end{table}


The LOS probability given in the literature (including the Winner model) usually refers to a LOS connection between {\em users}, not necessarily between the {\em antennas} on the devices held by the users. In other words, there are situations where a transmit and receive antenna nominally have LOS (according to the model definition), because there are no {\em environmental obstacles} between them; however, the bodies of the users and/or the device casings might prevent an actual LOS. We will henceforth refer to this situation as ``body-obstructed LOS" (BLOS). It is especially critical at mm-wave frequencies, which to which the human body is essentially impervious. For microwaves, the human body can be taken into account by introducing an additional shadowing term.  

Let us first consider the case of mm-wave propagation. Since smartphones are usually held close to the chest, approximately half the cases of ``nominal" LOS are actually BLOS, while the rest is ``true" LOS: 
\be
P_{{\rm BLOS}} = P_{{\rm LOS}} = \frac{1}{2} \cdot P^{W}_{{\rm LOS}} ,
\ee
For the case of BLOS, alternative propagation paths such as reflections by walls, can sustain links, but the resulting path loss and related parameters are different from the ``true" LOS; thus separate parameterization has to be used. The case of non-line-of-sight (NLOS)\footnote{One example of NLOS transmission is that the transmitter and the receiver are in different rooms, with walls between them.}, clearly occurs with probability $1- P^{W}_{{\rm LOS}}$. In the case of mm-wave communications, walls constitute an insurmountable obstacle, i.e., penetration of radiation into neighboring rooms, and between inside/outside the building, is negligible. 

For microwave propagation, the effect of body shadowing is better explained by an additional lognormal fading. In contrast to the ``standard" shadowing that describes shadowing by environmental obstacles and that changes as users move laterally, body shadowing variations are created by rotation of the users - resulting in the highest attenuation when they are standing back-to-back.  In \cite{karedal2008measurement}, it is shown that the body shadowing attenuation $\chi_{\sigma_{Lb}}$ follows log-normal distribution with mean $0$ and standard deviation ${\sigma_{Lb}}$. For D2D communication, we use the hand-to-hand model (HH2HH) as shown in \cite{karedal2008measurement}.

\subsubsection{Device-to-device channels at $38$GHz}

I this case, the pathloss is given by 
\be
\label{eq: channel model 3}
PL(d) = 20\log_{10}\left(\frac{4\pi d_0}{\lambda}\right) + 10\alpha\log_{10}\left(\frac{d}{d_0}\right) + \chi_{\sigma},
\ee
where 
$d_0 = 5$m is the free-space reference distance, 
$\lambda$ is the wavelength, $\alpha$ is the average pathloss, $\chi_{\sigma}$ is the shadowing parameter with mean $0$ and standard deviation $\sigma$. We assume that no $38$ Hz communication is possible when $d > 80$ m. From \cite{rappaport2011broadband} \cite{joonICC2013}, the system parameters are given by: $\alpha_{LOS} = 2.21$, $\alpha_{NLOS} = 3.18$, $\sigma_{LOS} = 9.4$ and $\sigma_{NLOS} = 11$.

\subsubsection{Device-to-device channels at $2.4$GHz}

For this case we can directly use the Winner II channel model, although we assume that no communication is possible for a distance larger than $100$m.\footnote{This is a conservative assumption motivated by the fact that at low SNR it is difficult for a D2D link to acquire beacon signals and discover other D2D devices.} Since $2.4$ GHz communication can penetrate walls, we have to account for difference scenarios, which are indoor communication (Winner model A1), outdoor-to-indoor communication (B4), indoor-to-outdoor communication (A2), and outdoor communication (B1).

We illustrate the case of the indoor (A1) communication, where the path loss model for both LOS and NLOS is given by \cite{winner2007d1},
\be
\label{eq: channel model 6}
PL(d) = A_1\log_{10}(d) + A_2 + A_3\log_{10}(f_c[GHz]/5) + X + \chi_{\sigma} ,
\ee
where $f_c$ is the carrier frequency. $A_1$ includes the path loss exponent. $A_2$ is the intercept and $A_3$ describes the path loss frequency dependence. $X = 5n_w$ is the (light) wall attenuation parameter, where $n_w$ is the number of walls between transmitter and receiver. $\chi_\sigma$ is the shadowing parameter assumed to be a log-normal distribution with mean $0$ and standard deviation $\sigma$, where $\sigma_{\rm LOS} = 3$ and $\sigma_{\rm NLOS} = 6$. Note that according to our discussion above, we add the body shadowing loss to Eq. (\ref{eq: channel model 6}), where for LOS, $\sigma_{L_{b}} = 4.2$ and for NLOS, $\sigma_{L_{b}} = 3.6$. All the other parameters for the indoor pathloss channel model in $2.4$ GHz are summarized in Table \ref{table: channel 4}. 

\begin{table}[ht]
\caption{The channel parameters for $2.4$ GHz D2D communications}
\centering 
\begin{tabular}{c | c | c | c | c | c | c  } 
\hline
\hline 
& $A_1 (LOS)$ & $A_2 (LOS)$ & $A_3 (LOS)$ & $A_1 (NLOS)$ & $A_2 (NLOS)$ & $A_3 (NLOS)$ \\ [0.5ex] 
\hline 
 indoor office & $18.7$ & $46.8$ & $20$ & $36.8$ & $43.8$ & $20$ \\ 
 \hline
 indoor hotspot & $13.9$ & $64.4$ & $20$ & $37.8$ & $36.5$ & $23$ \\
[1ex] 
\hline
\end{tabular}
\label{table: channel 4} 
\end{table}

For the other three cases, namely outdoor (B1), indoor-to-outdoor (A2) and outdoor-to-indoor (B4), we similarly directly use the respective Winner II channel models with antenna heights of $1.5$m, probabilistic LOS, and with the consideration of body shadowing.

\subsubsection{Channel between the Base Station and Devices}

In this case the Winner II channel model can also be used directly. In particular we use the urban macro-cell (C2) model for outdoor to outdoor communications and the urban macro outdoor to indoor (C4) model for outdoor to indoor communication; the only modification is the addition of the rotational body shadowing $\chi_{\sigma_{L_b}}$. As model for the rotational body shadowing, we use the access point to handheld device model (AP2HH \cite{karedal2008measurement}: for the case of LOS, $\sigma_{L_b}=2.3$ dB, while for NLOS, it is $\sigma_{L_b}=2.2$ dB.) 


For example, for the urban macro-cell (C2) channel model, the pathloss for LOS of the Winner model (i.e., without body shadowing) is given by
\begin{align}
PL_{\rm LOS}(d) = \left\{\begin{array}{cc} A_1\log_{10}(d) + A_2 + A_3\log_{10}(f_c[{\rm GHz}]/5) + \chi_{\sigma_1} , & 10 {\rm m} < d < d_{\rm BP}' \\ 
40\log_{10}(d) + 13.37 - 14\log_{10}(h_{\rm BS}') , & d_{\rm BP}' < d < 5000 {\rm m} \\
- 14\log_{10}(h_{\rm MS}') + 6\log_{10}(f_c[{\rm GHz}]/5) + \chi_{\sigma_2} \end{array}\right.,
\end{align}
where  $d'_{\rm BP} = 4h_{\rm BS}'h_{\rm MS}' f_c/c$ and $h_{\rm BS}' = h_{\rm BS} - 1$ and $h_{\rm MS}'  = h_{\rm MS} - 1$. We pick $h_{\rm BS} = 25$m and $h_{\rm MS}' = 1.5$m. $\chi_{\sigma_1}$ and $\chi_{\sigma_2}$ are shadowing attenuations, which are lognormally distributed with mean $0$ and standard deviation $\sigma_1 = 4$ and $\sigma_2 = 6$.
For NLOS, we have
\begin{eqnarray}
& & PL_{\rm NLOS}(d) \notag\\ 
& & = (44.9 - 6.55\log_{10}(h_{\rm BS}))\log_{10}(d) + 34.46 + 5.83\log_{10}(h_{\rm BS}) + 23\log_{10}(f_c[{\rm GHz}]/5) + \chi_{\sigma}, \notag\\
\end{eqnarray}
where $50 {\rm m} < d < 5000 {\rm m}$. The shadowing $\chi_{\sigma}$ is zero-mean and has standard deviation $\sigma = 8$. 
Similarly, the urban outdoor to indoor (C4) channel model can be found in \cite{winner2007d1}.

Moreover, to simulate the realistic scenario, we also assume a frequency reuse factor $K$ in this case to avoid the interference between cells \cite{molisch2011wireless}. 



\subsubsection{Link Capacity Computation}
By given all the system parameters, the link capacity for a transmitter-receiver pair is given by
\be
\label{eq: channel model 1}
C = B \cdot \log_2(1+{\rm SINR})
\ee
where ${\rm SINR} = P_{\rm signal}/(P_{\rm noise} + P_{\rm interference})$ (Signal to Interference plus Noise Ratio), and $B$ denotes the signal channel bandwidth. 
Specifically, on a dB scale, $P_{\rm signal}$ is given by
\be
\label{eq: channel model 2}
P_{\rm signal, dB} = P_{TX} + G_t + G_r - PL(d) 
\ee
where the $P_{TX}$ is the transmit power, $G_t$ and $G_r$ are the transmit and receive antenna gains. 
$P_{\rm interference}$ is the sum of the all the interference to a receiver.\footnote{The  model for mm-wave communication is considered to be interference free ($P_{\rm interference}=0$) since the angle of arrival (AOA) is very narrow (less than 10 degree).}

On a dB scale, the noise power is given by 
\be
\label{eq: channel model 5}
P_{\rm noise, dB} = 10\log_{10}(k_BT_e) + 10\log_{10}B + F_N, 
\ee
where $k_BT_e = -174$ dBm/Hz is the noise power spectral density and $F_N=6$ dB is a typical noise figure of the receiver. We assume this model to hold at all frequencies.\footnote{While for the same cost, receivers at 2 GHz might provide a better noise figure due to better-established fabrication processes, the impact of this effect on the system performance is low, and will be neglected henceforth.} The parameters of the three types of transmissions are summarized in Table \ref{table: channel 2}.

\begin{table}[ht]
\caption{The parameters for the three types of transmissions }
\centering 
\begin{tabular}{c | c | c | c | c | c | c} 
\hline
\hline 
& $B$ & $f_c$ & $P_{TX}$  & $G_t$ & $G_r$ & $K$ \\ [0.5ex] 
\hline 
mm-wave D2D transmissions & $800$ MHz & $38$ GHz & $20$ dBm & $9$ dB & $9$ dB  & $4$\\ 
\hline 
microwave D2D transmissions & $20$ MHz & $2.45$ GHz or $2.1$ GHz & $20$ dBm  & $12$ dB & $0$ & $4$\\ 
\hline 
cellular transmissions & $20$ MHz & $2.1$ GHz & $43$ dBm & $12$ dB & $0$ & $3$ \\ 
[1ex] 
\hline 
\end{tabular}
\label{table: channel 2} 
\end{table}


\subsection{Results and Discussions}
In this section, we will present the simulation results. If not stated otherwise, we will use the following settings: the number of users is $n=10000$; the users are uniformly and indepdently distributed in the cell 
(We can show that a negligible difference between regular grid and random distribution by simulation (not shown here); we thus henceforth show only results for the random node distribution). 
The number of files in the library is $m=300$, which is representative of the library size 
of a video on demand service.\footnote{In practice, the library of titles in such a service would be refreshed every few days.} The user cache size is $M = 20$ files unless specifically mentioned, which even with high definition (HD) quality requires less than the (nowadays) ubiquitous $64$ GByte of storage space. We let each user independently make a request by sampling from 
a Zipf distribution with $\gamma_r = 0.4$; this value is at the lower edge of the range of values that have been measured in practice \cite{breslau1999web}; note that the advantages of caching would be {\em more} pronounced for larger $\gamma_r$. The interference between concurrent D2D links sharing the same frequency band is treated as noise. For the harmonic broadcasting, we chose a video file size of $L=2.7$ Gbits and the number of blocks $P=540$ \cite{sanchez2011idash}.

\subsubsection{Throughput-Outage Tradeoff}

\begin{figure}
\centering
\includegraphics[width=11cm]{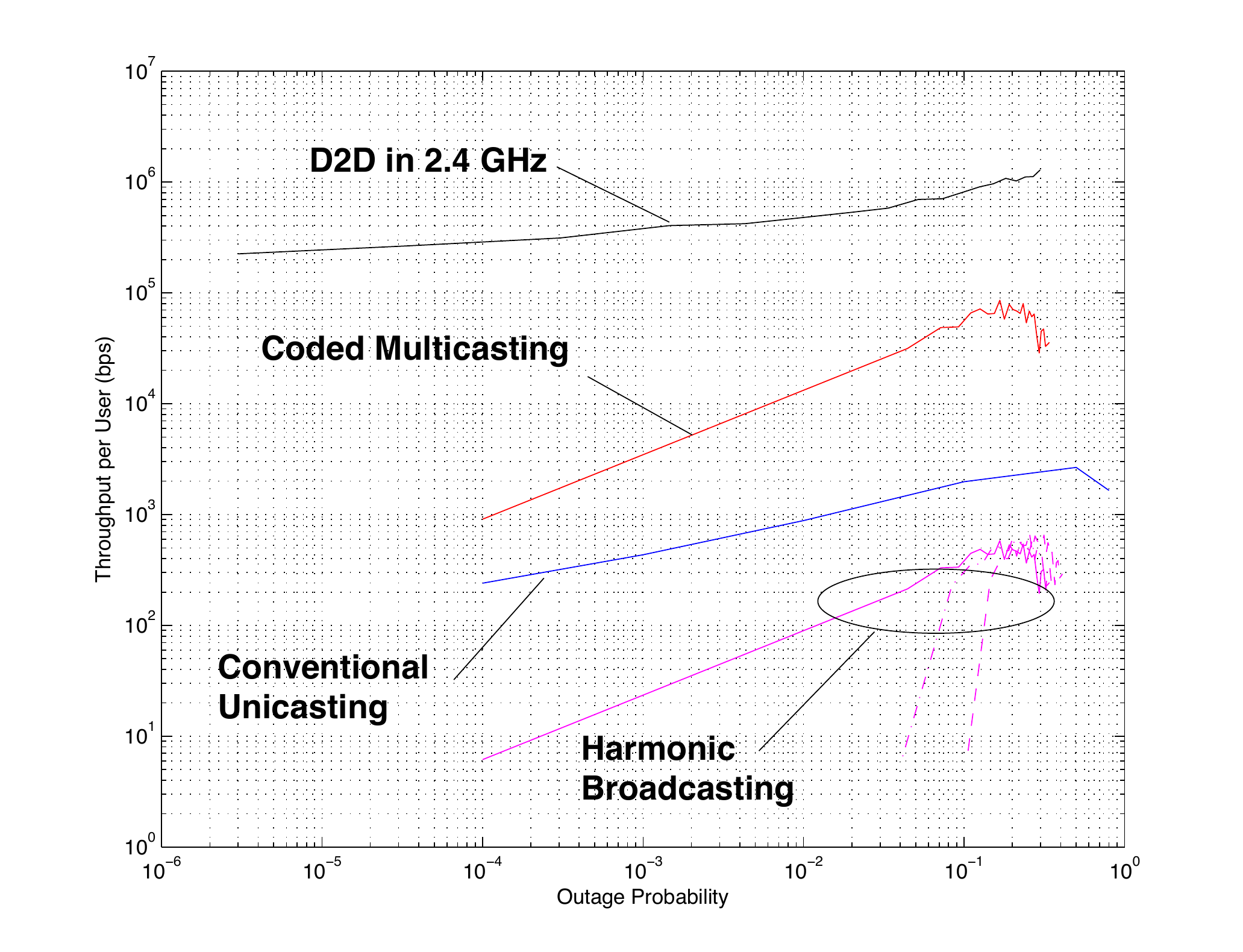} 
\caption{Simulation results for the throughput-outage tradeoff for conventional unicasting, coded multicasting, harmonic broadcasting and the $2.45$ GHz D2D communication scheme under both indoor office and indoor hotspot channel models. Solid lines: indoor office; dashed lines: indoor hotspot.}
\label{fig: result_1}
\end{figure}

In Fig. \ref{fig: result_1}, we plot the performance of all the discussed schemes separately, where a $2.45$ GHz D2D only scheme is implemented. From Fig. \ref{fig: result_1}, we can see that the throughput of the D2D scheme is markedly (order of magnitude at low outages) higher than the conventional unicasting, harmonic broadcasting and even coded multicast scheme. This shows that in practical situations, the ``scaling law" is not the only aspect of importance. Rather, the higher capacity of the short-distance links plays a significant role, and a good throughput-outage tradeoff can be achieved even without the use of a BS connection as ``backstop". The main reason lies in the fact that for the coded mulicasting or harmonic broadcasting scheme, outage is determined by bad channel conditions, and no diversity is built into the system. For D2D, even though the outage in our scheme are caused by both physical channel and the lack of the requested files in the corresponding cluster, the channel diversity plays a more importance role. Moreover, Fig. \ref{fig: result_1} shows that the behaviors of all schemes hold for both the indoor office and the indoor hotspot environment. 

\begin{figure}
\centering
\includegraphics[width= 11cm]{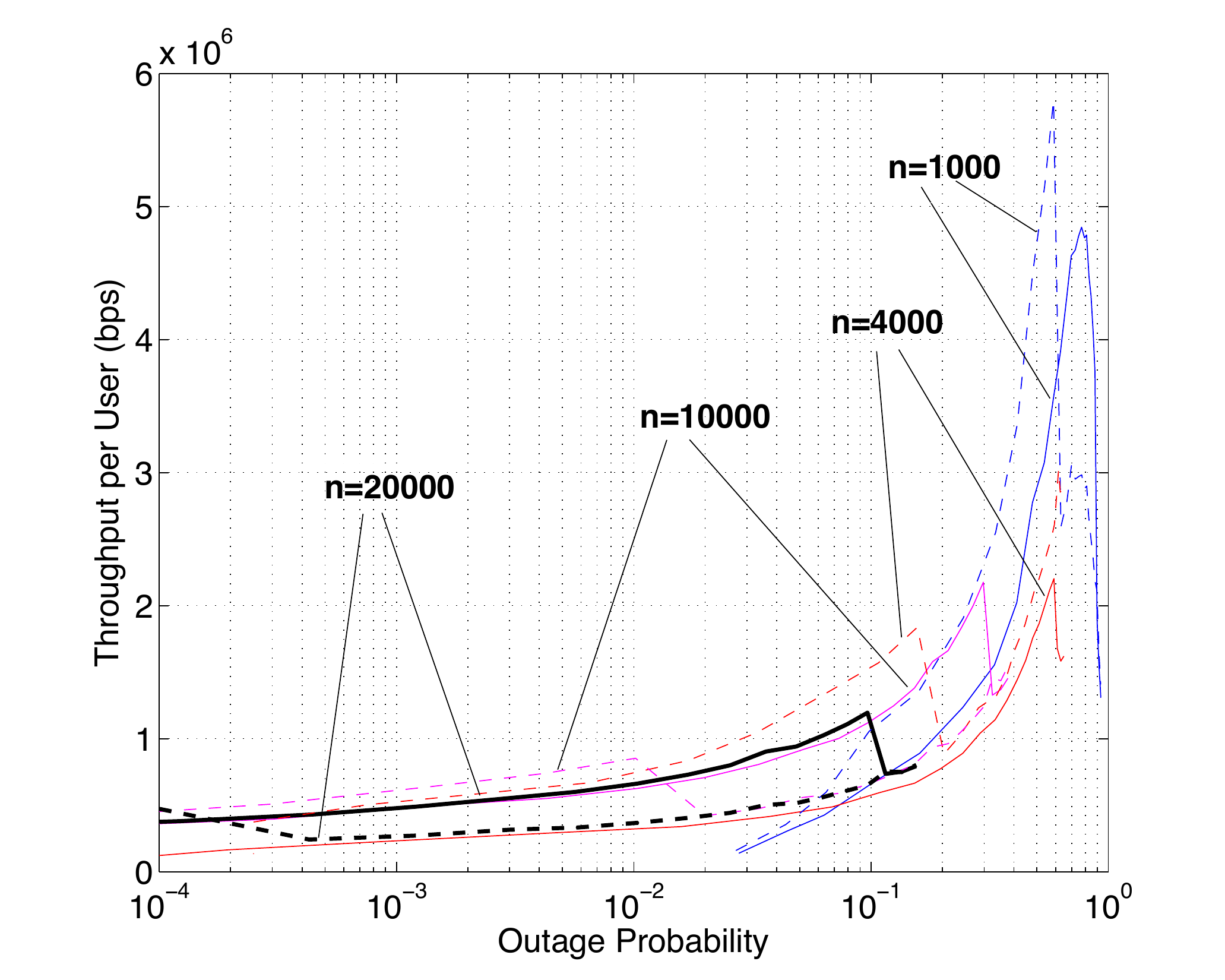} 
\caption{The throughput-outage tradeoff for different user densities. Solid lines: indoor office; dashed lines: indoor hotspot. }
\label{fig: result_6}
\end{figure}

In Fig.~\ref{fig: result_6}, for the hotspot, we furthermore obtain the interesting result that the throughput-outage tradeoff is non-monotonous if we use the (theoretically derived) cluster size. 
This behavior is caused by a higher LOS probability when the cluster size becomes small: there is an appreciable probability that the useful signal is NLOS but there exist some LOS interferers. From Fig.~\ref{fig: result_6}, \ref{fig: result_7} and \ref{fig: result_11}, the similar phenomena can also be observed for the case of the indoor office model but for different parameter settings. Of course this does not mean that the optimum throughput-outage tradeoff in practice is non-monotonous; rather it is a consequence of using a cluster size derived under one specific model in the deployment by using a different model. 

In Fig. \ref{fig: result_1}, the fact that our D2D scheme performs better in the hotspot scenario than in the indoor office case is mainly due to the low probability of LOS from interferers and the high probability of LOS for useful signal  (note that the LOS probability in the hotspot is unity up to distances of 10 m and decreases exponentially for larger distances). However, for the coded multicast transmission, the performance in indoor hotspot is actually worse than the indoor office model; this is due to the larger size of the buildings so that the pathloss caused by $d_{\rm in}$ in the urban macro outdoor-to-indoor model (C4) \cite{winner2007d1} is very significant, where $d_{\rm in}$ is the distance from the wall to the indoor terminal.

Besides the performance advantage of the D2D approach (compared to coded multicast), it also has the advantage of a simpler caching placement and delivery. The coded multicasting approach in \cite{maddah2012fundamental} constructs the cache contents and the coded delivery scheme in a combinatorial manner which does not scale well with $n$. For example, in our network configuration, it requires a code length larger than ${10000 \choose 600}$, which is larger than $10^{15}$.

\subsubsection{Holistic Multi-Frequency D2D System Performance}

In this section, we investigate the performance of the proposed D2D system given by Fig. \ref{fig: FlowChart} in Section \ref{sec: holistic desigen}. 

\begin{figure}
\centering
\includegraphics[width=11cm]{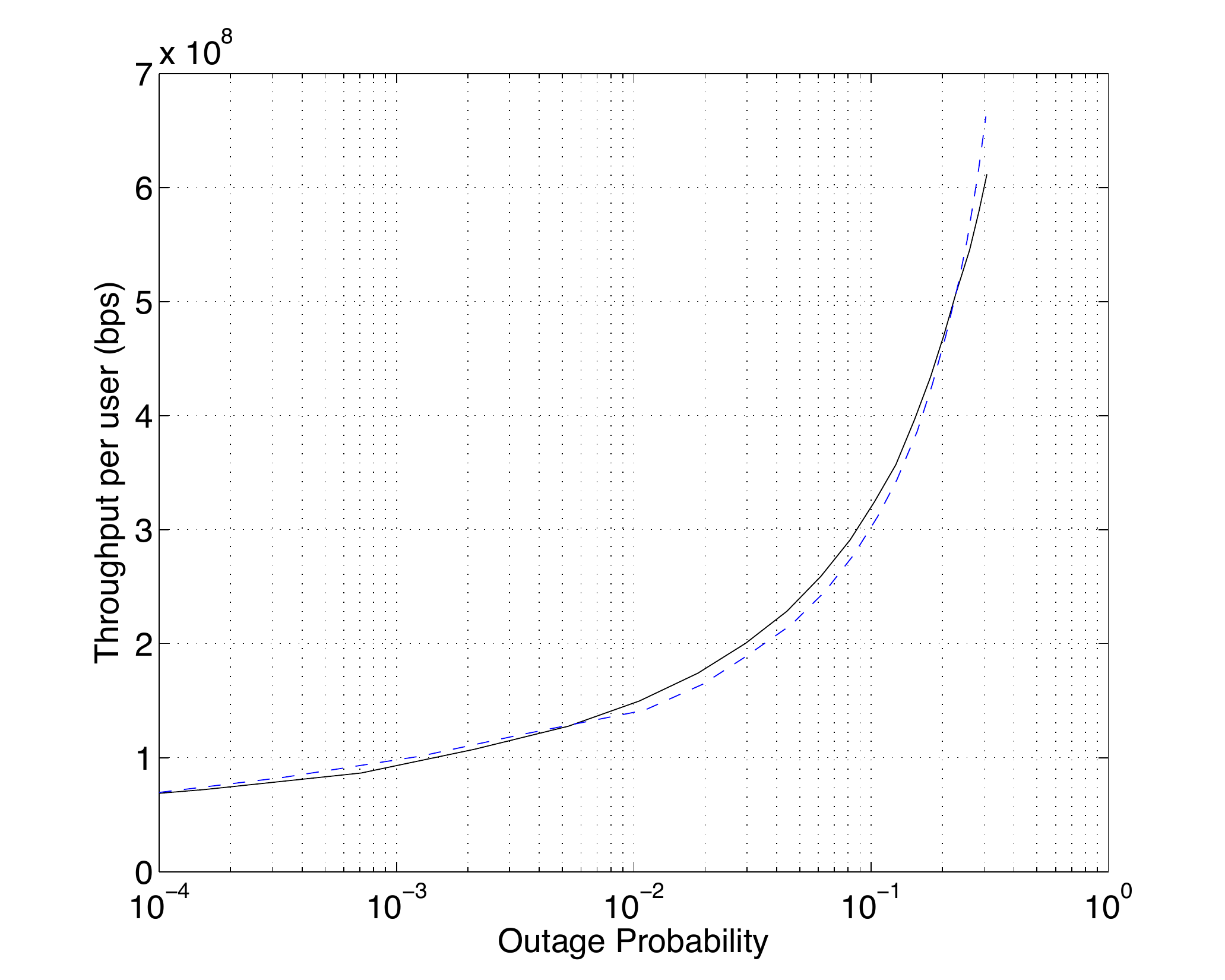} 
\caption{Simulation results for the throughput-outage tradeoff by holistic system design. Black Solid lines: indoor office; blue dashed lines: indoor hotspot.}
\label{fig: result_4}
\end{figure}

\begin{figure}
\centering
\includegraphics[width=11cm]{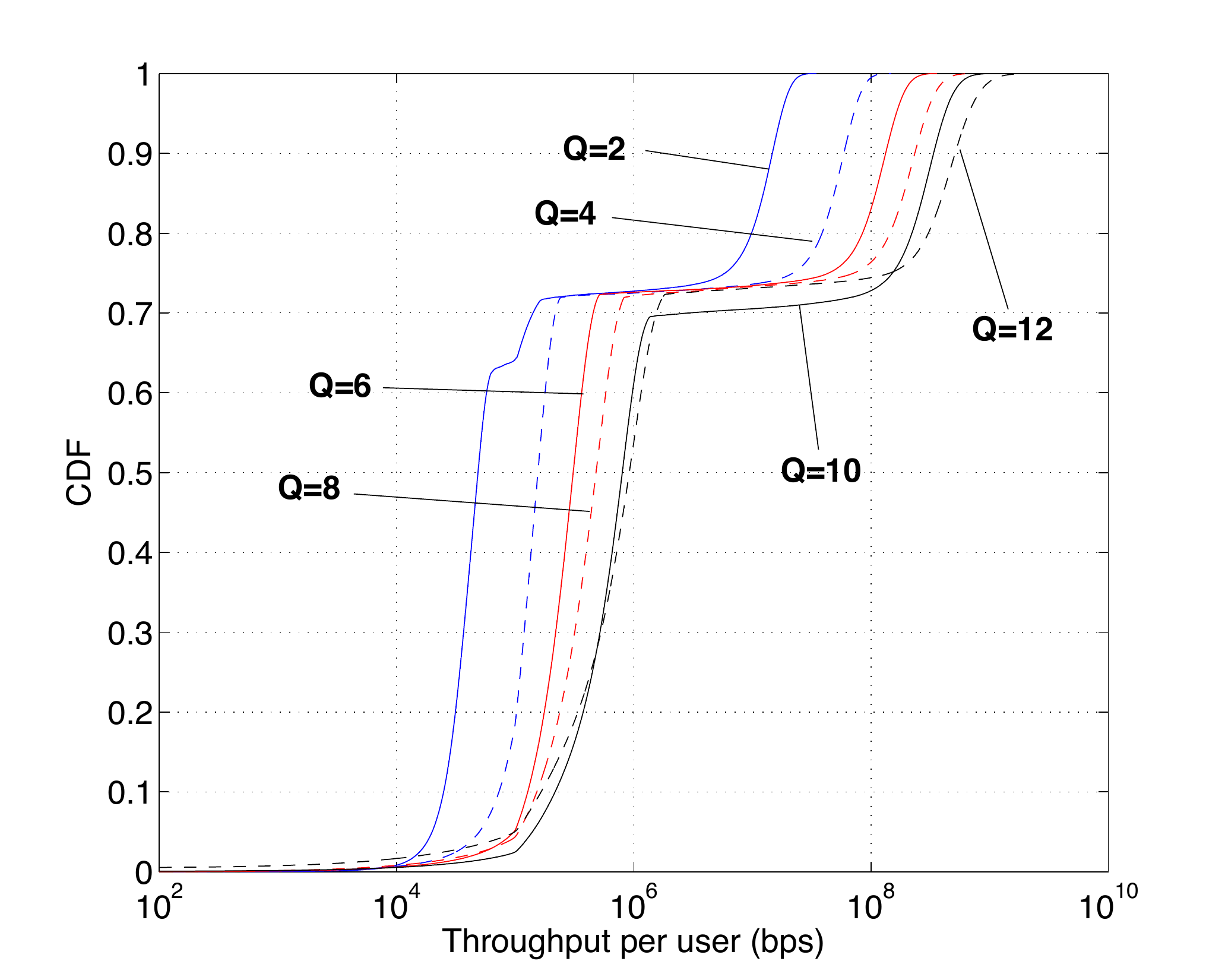} 
\caption{The CDF of the throughput for different outage probabilities (cluster size of $\frac{600^2}{Q^2}$) under indoor office model.}
\label{fig: result_2}
\end{figure}

Fig. \ref{fig: result_4} shows that the average throughput per user increases significantly due to the help of the $38$ GHz D2D communications. Consider the CDF (Cumulative distribution function) of the throughput for different outage probabilities shown in Fig.~\ref{fig: result_2}, in this way, we can see on average, how many users will be served with a throughput that is less than the minimum required rate for video streaming, for example, $100$Kbps. 

\begin{figure}
\centering
\includegraphics[width=11cm]{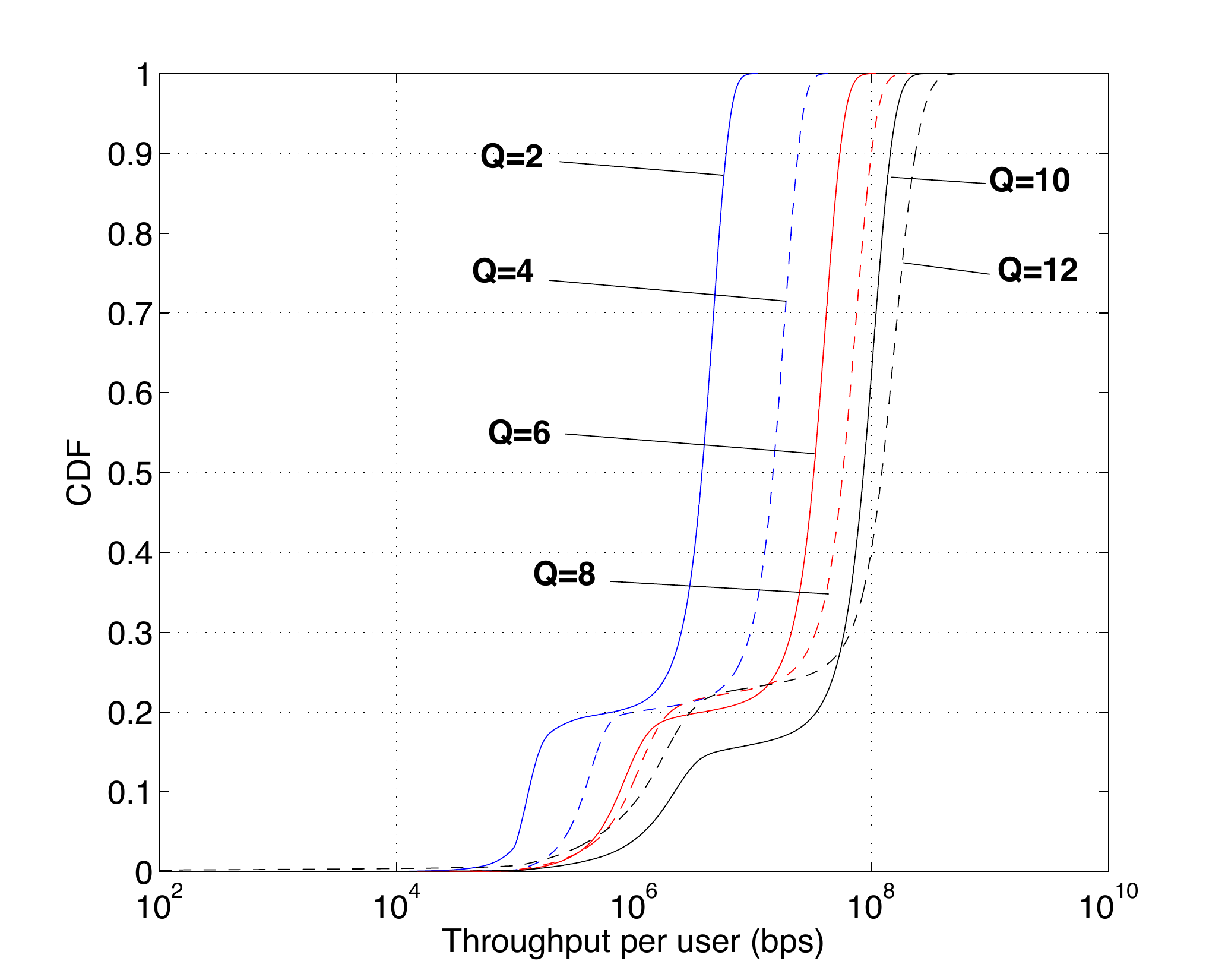}  
\caption{The CDF of the throughput for different outage probabilities (cluster size of $\frac{600^2}{Q^2}$) under indoor hotspot model.}
\label{fig: result_3}
\end{figure}

Fig.~\ref{fig: result_2} and \ref{fig: result_3} show the throughput as a function of the cluster size. Intuitively, a large cluster size corresponds to a small outage probability, as the probability is high that the desired file is found within a cluster. This is reflected by the throughput CDF: a small cluster size results in a small minimum throughput, but a large maximum throughput (compare, e.g., the red-dotted line in Fig.~\ref{fig: result_3}; this is similar to the effect we have observed in the previous subsection. For example, if we pick a cluster size of $100 {\rm m} \times 100 {\rm m}$, then the number of users whose rate is less than  $100$ Kbps only around $250$. Moreover, about $30\%$ of users are served with a data rate larger than $2$ Mbps) due to help by $38$ GHz D2D communications.

\begin{figure}
\centering
\includegraphics[width=11cm]{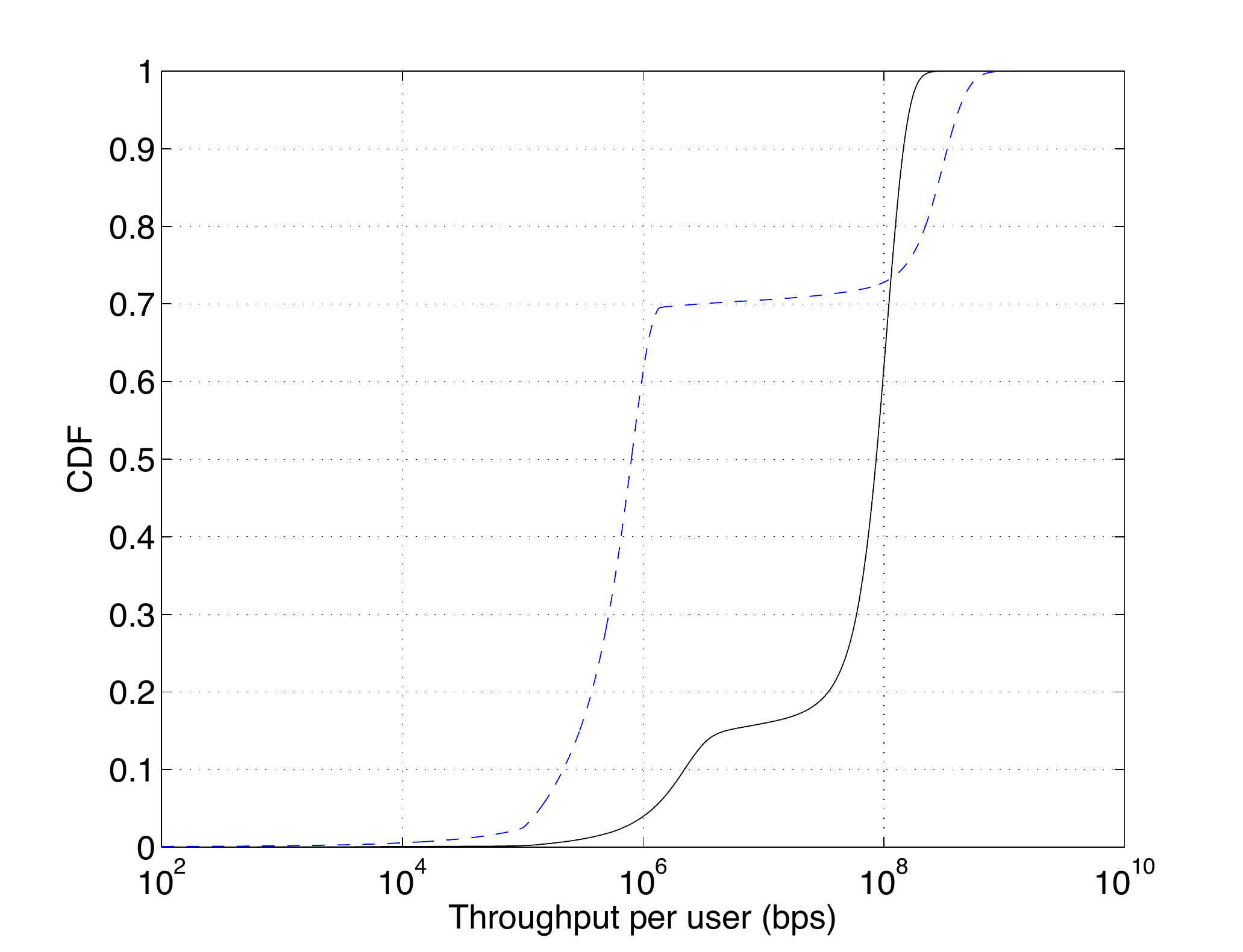} 
\caption{The CDF of the throughput for the cluster size of $100 {\rm m} \times 100 {\rm m}$ under indoor office and indoor hotspot channel model. Solid lines: indoor office; dashed lines: indoor hotspot.}
\label{fig: result_5}
\end{figure}

From Fig.~\ref{fig: result_4}, \ref{fig: result_2} and \ref{fig: result_3}, we observe similar behavior under the indoor hotspot model, where interestingly, the performance of our holistic multi-frequency design in terms of average throughput is not very different from that of the indoor office model. The reason is that the variance of the throughput for the indoor model is much larger than that for the indoor hotspot model, which is due to the fact that fewer users can be served by the $38$ GHz D2D communications.\footnote{We serve the users in a round robin fashion in one cluster even for $38$ GHz communications to avoid interference.} 
Both CDFs for the case of  $100 {\rm m} \times 100 {\rm m}$ cluster size are shown in Fig. \ref{fig: result_5}, 
In this case, almost no users have a service rate less than $100$ Kbps and about $90\%$ of users can get HD quality services. Moreover, we notice that the role of base station in this scenario is to reduce the outage probability. For example, when cluster size is $100 {\rm m} \times 100 {\rm m}$, the base station can serve $400 \sim 500$ users in the indoor office model. 

\subsubsection{Effects of the Density of Nodes}

From Theorem \ref{theorem: 4}, we expect that the throughput-outage tradeoff does not depend on the number of users or user density as long as $n$ and $m$ are large and $Mn \gg m$. 
However, the throughput-outage scaling behavior was obtained under the simplified protocol model, where the relation between the link rate and the link range (source-destination distance) is not specified. In practice, if we want to obtain a high communication rate, the D2D communication distance cannot be very large due to the large pathloss, which reduces the per-link capacity. This is especially true for $38$ GHz communications under the indoor office environment. Therefore, the user density is also an important parameter for the system performance. In this section, we investigate the system behavior for different user densities by focusing on the case of $2.45$ GHz D2D communications only.

In Fig.~\ref{fig: result_6}, we observe that there exists a tradeoff between the user density and the throughput, which is because that the impact of the user density on the link rate is twofold: on one hand, a higher user density allows a smaller cluster size, in turn resulting in shorter links and higher SINR. On the other hand, a small cluster size increases the probability for having LOS interference, which can degrade the performance of the system significantly. 

\subsubsection{Effects of the Storage Capacity and the Library Size}

\begin{figure}
\centering
\includegraphics[width=11cm]{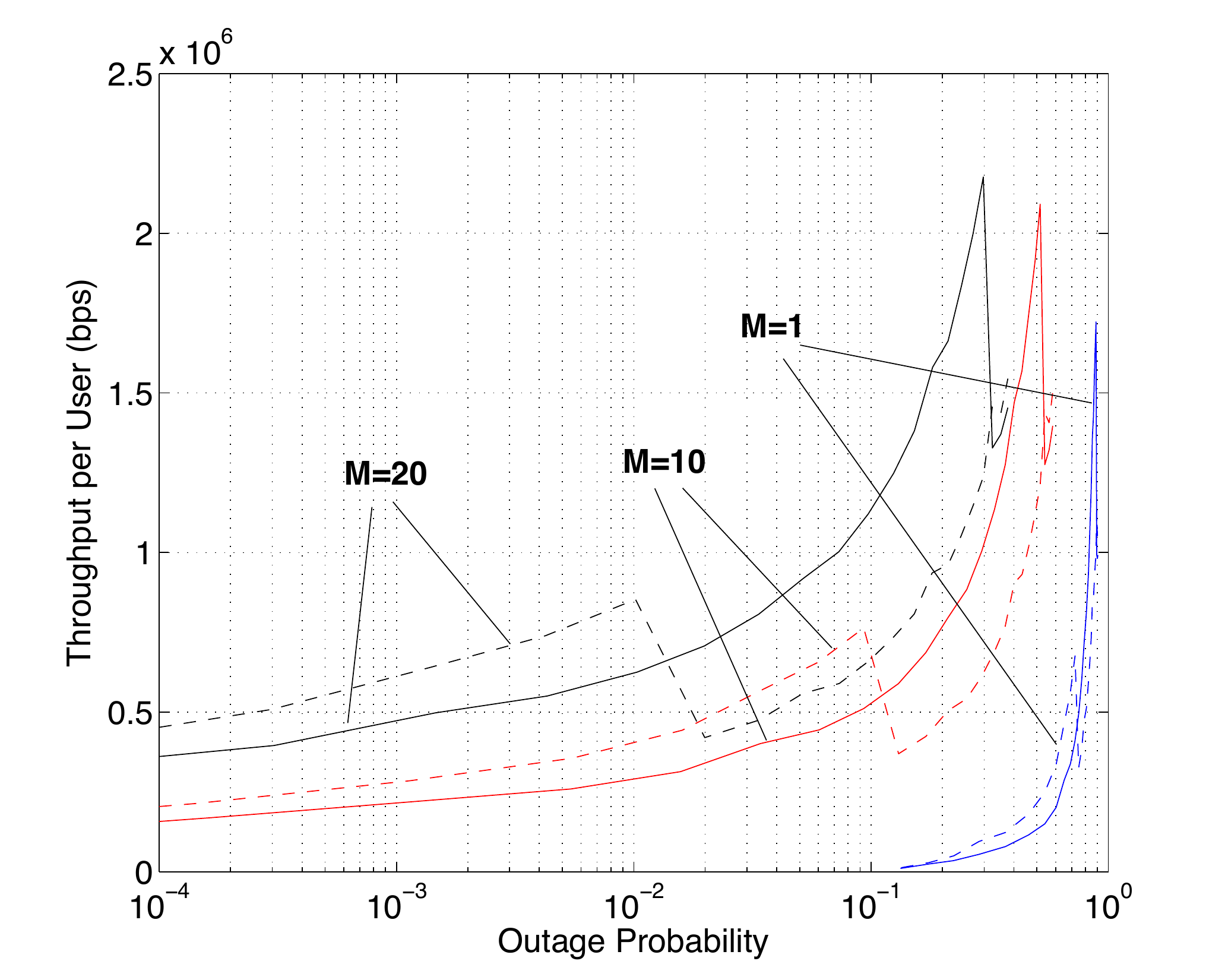} 
\caption{The throughput-outage tradeoff for different user storage capcity. Solid lines: indoor office; dashed lines: indoor hotspot. }
\label{fig: result_7}
\end{figure}

As already observed in Section \ref{sec: Key results}, in the regime $nM \gg m$ the D2D system yields a linear dependence of the throughput on the user storage capacity $M$. This means that such a system can directly trade cache memory for throughput. Since storage memory is a cheap and rapidly growing network resource, while bandwidth is scarce and very expensive, the attractiveness of the D2D approach is self-evident. 
The result also holds true in practice, as demonstrated by the simulation results in Fig. \ref{fig: result_7}. We observe that, when the outage is small, the average throughput per user increases even faster than linearly with $M$. This is because in practice the link rate $C_r$ is a decreasing function of the link range. Therefore, when $M$ becomes large, we can decrease the D2D cluster size (and therefore the average link range) while maintaining a constant outage probability.

\begin{figure}
\centering
\includegraphics[width=11cm]{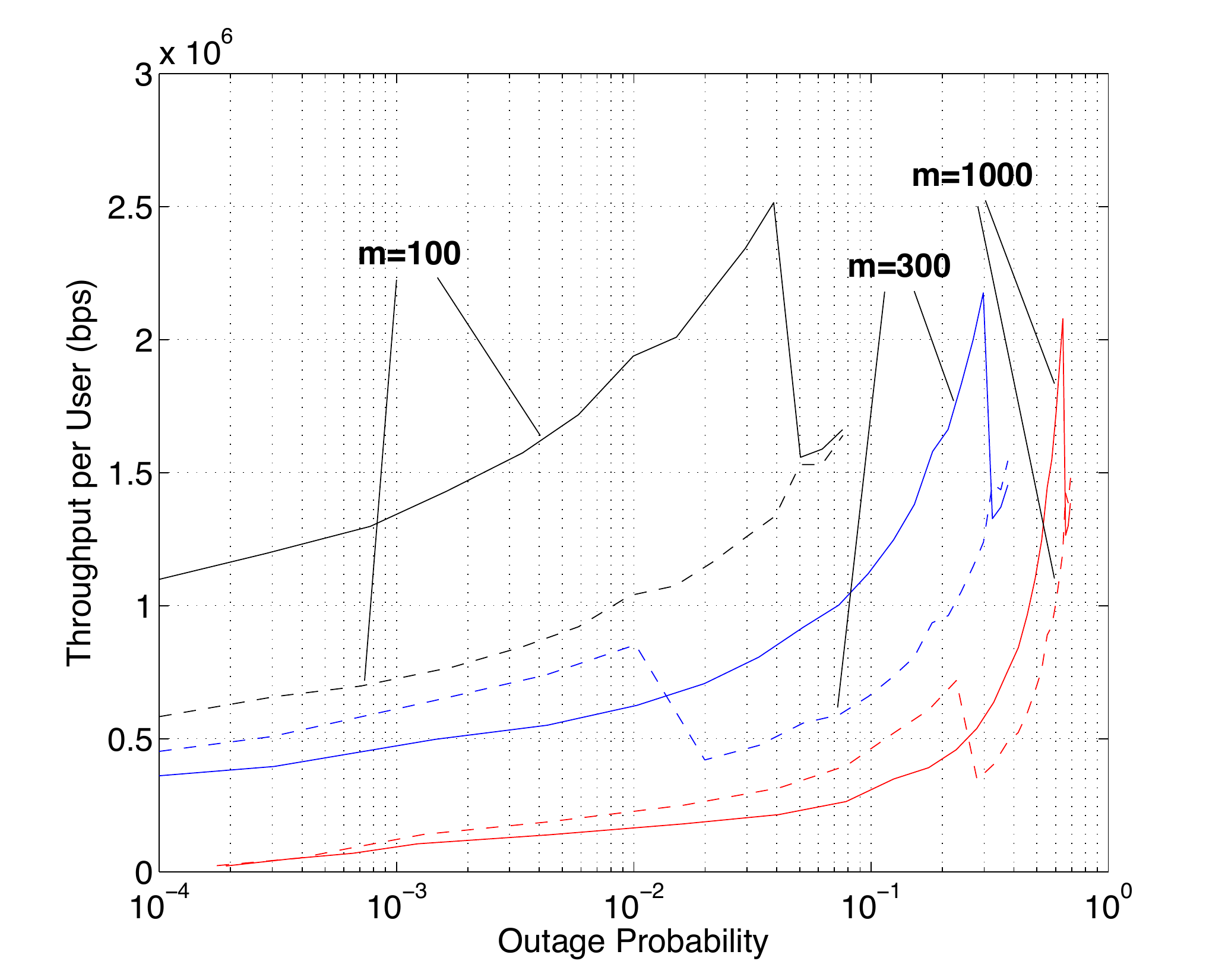} 
\caption{The throughput-outage tradeoff for different library size of files. Solid lines: indoor office; dashed lines: indoor hotspot. }
\label{fig: result_11}
\end{figure}

Fig.~\ref{fig: result_11} shows the throughput-outage tradeoff for different library size. As expected from Theorem \ref{theorem: 4}, 
in this case we notice that the throughput decreases roughly inversely proportional to the library size $m$, for fixed cache capacity $M$.

\subsubsection{$2.1$ GHz In Band Communications}

\begin{figure}
\centering
\includegraphics[width=11cm]{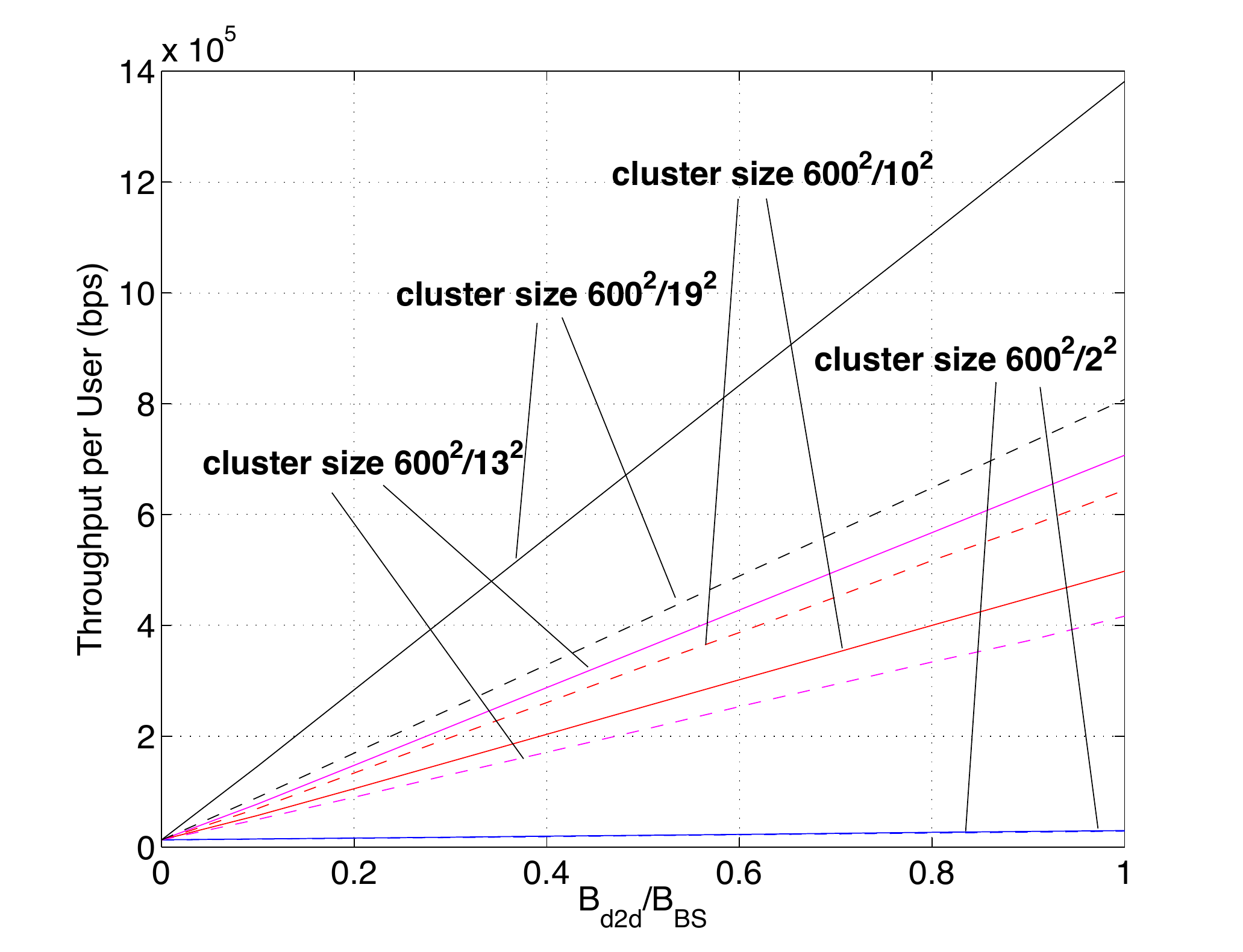} 
\caption{The throughput v.s. bandwidth division between $2.1$ GHz communication and the base station under different cluster size, where $B_{\rm d2d}$ is the bandwidth by $2.1$ GHz communications and $B_{\rm BS}$ is the bandwidth by the cellular base station. $B_{\rm d2d} + B_{\rm BS} = B = 20$MHz. Solid lines: indoor office; dashed lines: indoor hotspot. }
\label{fig: result_8}
\end{figure}

Sometimes, the D2D communications and the cellular communications by the BS have to share the same spectrum, which raises the question of how to divide the bandwidth for each type of communications. Obviously, this depends on the channel realizations for each type of communications. From our simulations, we obtain that under our channel model (either indoor office or indoor hotspot), the base station cannot support more than about $1000$ users in the best case if the minimum video coding rate is $100$ Kbps, while, for $2.1$ GHz D2D communication, it is very easy to support a much larger number of users at a certain playback rate requirement.  Therefore, it is intuitive that there is no tradeoff between the bandwidth division and the {\em average} throughput by fixing the cluster size (outage probability in Theorem \ref{theorem: 4}). The simulation results are shown in Fig. \ref{fig: result_8}, which confirm our intuition. 

On the other hand, if we care more about the outage probability, then there is a clear tradeoff between the outage probability and the division of the bandwidth, especially for the small cluster size. This occurs because the BS is capable of satisfying ``costly" links that normally would either increase outage probability or would enforce an increase in cluster size. 

\begin{figure}
\centering
\includegraphics[width=11cm]{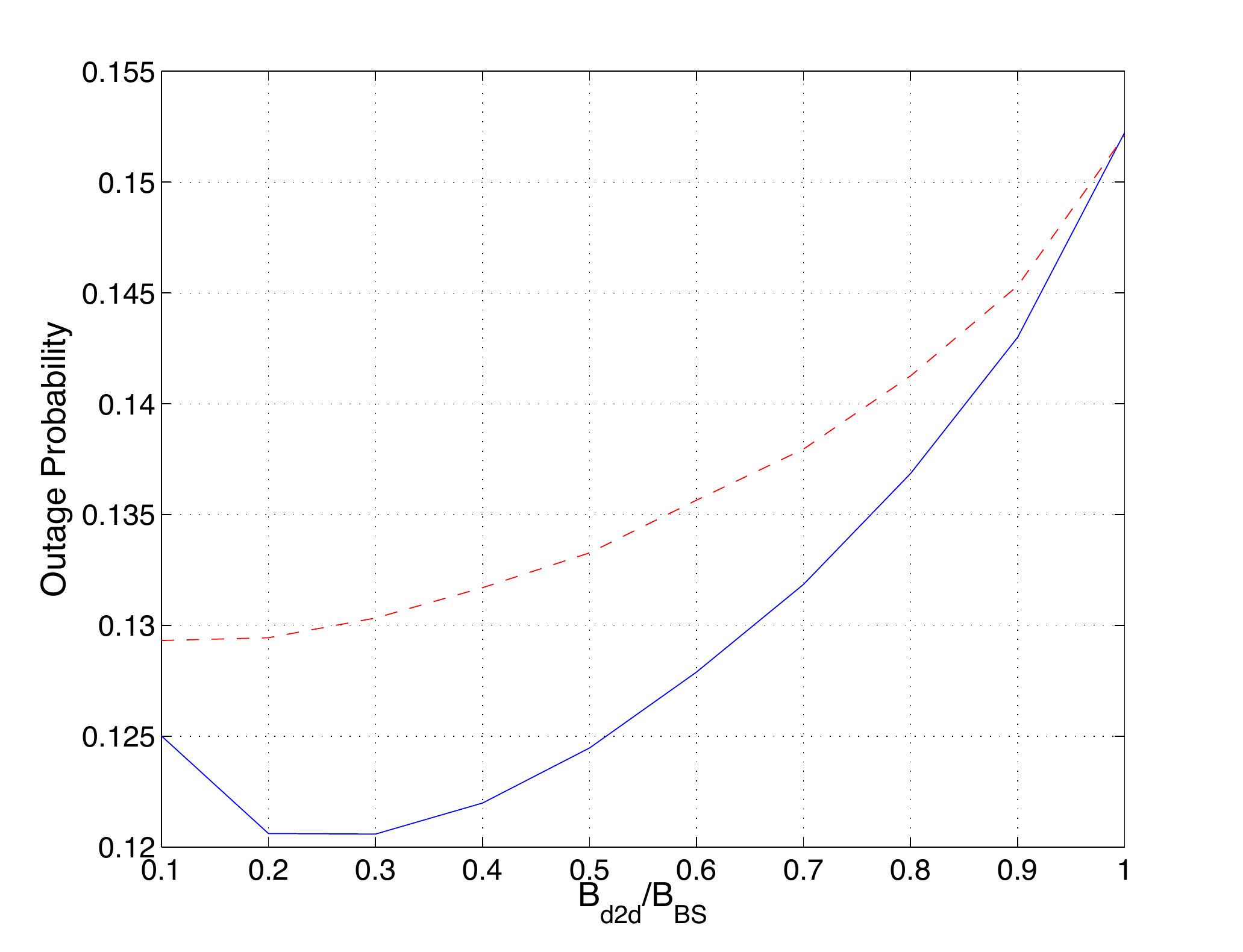} 
\caption{The outage v.s. bandwidth division between $2.1$ GHz communication and the base station for the cluster with size $600^2/19^2$. Blue solid lines: indoor office; red dashed lines: indoor hotspot.}
\label{fig: result_10}
\end{figure}

In Fig.~\ref{fig: result_10}, under the office channel model, when the area of the cluster is $600^2/19^2$, the best bandwidth division is when $B_{\rm d2d}/B_{\rm BS} = 0.2$, which means that we need to only allocate $20\%$ of the bandwidth to the D2D communication to obtain the minimum outage probability. Similar behavior can also be observed for the hotspot model with the difference that now $B_{\rm d2d}/B_{\rm BS} = 0.1$ is the best bandwidth division, which is because that the link rate under the hotspot channel model is better than that for the indoor office model.

\section{Conclusions}
\label{sec: Conclusions}

In this paper we have reviewed in a concise and tutorial fashion some recent results on base station assisted D2D wireless networks with caching for video delivery, recently proposed by the authors \cite{ji2013optimal}, 
as well as some competing conventional schemes and a recently developed scheme based on caching at the user devices but involving only coded multicasting transmission from the base station. We have reviewed the throughput-outage scaling laws of such schemes on the basis of a simple protocol model which captures the fundamental aspects of interference and spatial spectrum reuse through geometric link conflict constraints. This model allows  a  sharp characterization of the throughout-outage tradeoff in the asymptotic regime of dense networks. This tradeoff shows the superiority of the D2D caching network approach and of the 
coded multicasting approach over the conventional schemes, which can be regarded as today current technology. 

In order to gain a better understanding of the actual performance in realistic environments, we have developed an accurate simulation model and some guidelines for the system design. In particular, we have considered a holistic system design including D2D links at $38$ GHz and $2.45$ (or $2.1$) GHz, and the cellular downlink at $2.1$ GHz, representative of an LTE network. 

We compared the schemes treated in the tutorial part of the paper on the basis of their throughput-outage tradeoff performance, and we have put in evidence several interesting aspects. 
In particular, we have shown the superiority of the D2D caching network even in realistic propagation conditions, including all the aspects that typically are expected to limit D2D communications, such as NLOS propagation, limited link range, environment shadowing
and human body shadowing. Remarkably, even in such realistic conditions, the D2D caching network shows very competitive performance with respect to the other schemes. In particular, the proposed system is able to efficiently trade the cache memory in the user devices for the system throughput. Since the former is a rapidly growing, cheap  and yet untapped network resource, while the latter is known to be scarce and very expensive, the interest in developing and deploying such D2D caching networks becomes evident. This fact is even more remarkable 
if we consider the fact that the D2D network requires simple decentralized caching and does not require any sophisticated network coding technique to share the files over the D2D links.

\bibliographystyle{IEEEbib}
\bibliography{references,dilip-ref}

\begin{thebibliography}{10}

\bibitem{cisco66}
``{http://www.cisco.com/en/US/solutions/collateral/ns341/ns525/ns537
  /ns705/ns827/white/paper/c11-520862.html.},''
\newblock .

\bibitem{annapureddy2010coordinated}
S.~Annapureddy, A.~Barbieri, S.~Geirhofer, S.~Mallik, and A.~Gorokhov,
\newblock ``Coordinated joint transmission in wwan,''
\newblock in {\em IEEE Communication Theory Workshop}, 2010.

\bibitem{irmer2011coordinated}
R.~Irmer, H.~Droste, P.~Marsch, M.~Grieger, G.~Fettweis, S.~Brueck, H-P Mayer,
  L.~Thiele, and V.~Jungnickel,
\newblock ``Coordinated multipoint: Concepts, performance, and field trial
  results,''
\newblock {\em Communications Magazine, IEEE}, vol. 49, no. 2, pp. 102--111,
  2011.

\bibitem{6476878}
J.G. Andrews,
\newblock ``Seven ways that hetnets are a cellular paradigm shift,''
\newblock {\em Communications Magazine, IEEE}, vol. 51, no. 3, pp. 136--144,
  2013.

\bibitem{sanchez2011improved}
Y.~Sanchez, T.~Schierl, C.~Hellge, T.~Wiegand, D.~Hong, D.~De~Vleeschauwer,
  W.~Van~Leekwijck, and Y.~Lelouedec,
\newblock ``Improved caching for {HTTP}-based video on demand using scalable
  video coding,''
\newblock in {\em Consumer Communications and Networking Conference (CCNC),
  2011 IEEE}. IEEE, 2011, pp. 595--599.

\bibitem{sanchez2011idash}
Y.~S{\'a}nchez de~la Fuente, T.~Schierl, C.~Hellge, T.~Wiegand, D.~Hong,
  D.~De~Vleeschauwer, W.~Van~Leekwijck, and Y.~Le~Lou{\'e}dec,
\newblock ``idash: improved dynamic adaptive streaming over http using scalable
  video coding,''
\newblock in {\em Proceedings of the second annual ACM conference on Multimedia
  systems}. ACM, 2011, pp. 257--264.

\bibitem{begen2011watching}
A.~Begen, T.~Akgul, and M.~Baugher,
\newblock ``Watching video over the web: {P}art 1: Streaming protocols,''
\newblock {\em Internet Computing, IEEE}, vol. 15, no. 2, pp. 54--63, 2011.

\bibitem{li2012three}
Y.~Li, E.~Soljanin, and P.~Spasojevi{\'c},
\newblock ``Three schemes for wireless coded broadcast to heterogeneous
  users,''
\newblock {\em Physical Communication}, 2012.

\bibitem{jakubczak2010softcast}
S.~Jakubczak and D.~Katabi,
\newblock ``Softcast: one-size-fits-all wireless video,''
\newblock in {\em ACM SIGCOMM Computer Communication Review}. ACM, 2010,
  vol.~40, pp. 449--450.

\bibitem{aditya2011flexcast}
S.~Aditya and S.~Katti,
\newblock ``Flexcast: graceful wireless video streaming,''
\newblock in {\em Proceedings of the 17th annual international conference on
  Mobile computing and networking}. ACM, 2011, pp. 277--288.

\bibitem{bursalioglu2011lossy}
O.~Y. Bursalioglu, M.~Fresia, G.~Caire, and H.~V. Poor,
\newblock ``Lossy multicasting over binary symmetric broadcast channels,''
\newblock {\em Signal Processing, IEEE Transactions on}, vol. 59, no. 8, pp.
  3915--3929, 2011.

\bibitem{cover2006elements}
T.~M Cover and J.~A Thomas,
\newblock {\em Elements of information theory, 2nd Edition},
\newblock Wiley-interscience, 2006.

\bibitem{shokrollahi2006raptor}
A.~Shokrollahi,
\newblock ``Raptor codes,''
\newblock {\em Information Theory, IEEE Transactions on}, vol. 52, no. 6, pp.
  2551--2567, 2006.

\bibitem{luby2006raptor}
M.~Luby, M.~Watson, T.~Gasiba, T.~Stockhammer, and W.~Xu,
\newblock ``Raptor codes for reliable download delivery in wireless broadcast
  systems,''
\newblock in {\em Consumer Communications and Networking Conference, 2006. CCNC
  2006. 3rd IEEE}. IEEE, 2006, vol.~1, pp. 192--197.

\bibitem{oyman2010toward}
O.~Oyman, J.~Foerster, Y.~Tcha, and S.~Lee,
\newblock ``Toward enhanced mobile video services over wimax and lte [wimax/lte
  update],''
\newblock {\em Communications Magazine, IEEE}, vol. 48, no. 8, pp. 68--76,
  2010.

\bibitem{keller2012microcast}
L.~Keller, A.~Le, B.~Cici, H.~Seferoglu, C.~Fragouli, and A.~Markopoulou,
\newblock ``Microcast: Cooperative video streaming on smartphones,''
\newblock in {\em ACM MobiSys}. ACM, 2012, pp. 57--70.

\bibitem{juhn1997harmonic}
L-S. Juhn and L-M. Tseng,
\newblock ``Harmonic broadcasting for video-on-demand service,''
\newblock {\em Broadcasting, IEEE Transactions on}, vol. 43, no. 3, pp.
  268--271, 1997.

\bibitem{paris1998efficient}
J-F. P{\^a}ris, S.~W. Carter, and D.E. Long,
\newblock ``Efficient broadcasting protocols for video on demand,''
\newblock in {\em MASCOTS}. IEEE, 1998, pp. 127--132.

\bibitem{engebretsen2006harmonic}
L.~Engebretsen and M.~Sudan,
\newblock ``Harmonic broadcasting is bandwidth-optimal assuming constant bit
  rate,''
\newblock {\em Networks}, vol. 47, no. 3, pp. 172--177, 2006.

\bibitem{paris1998low}
J-F P{\^a}ris, S.~W Carter, and D.E. Long,
\newblock ``A low bandwidth broadcasting protocol for video on demand,''
\newblock in {\em Computer Communications and Networks, 1998. Proceedings. 7th
  International Conference on}. IEEE, 1998, pp. 690--697.

\bibitem{DBLP:journals/corr/abs-1109-4179}
N.~Golrezaei, K.~Shanmugam, A.~G Dimakis, A.~F Molisch, and G.~Caire,
\newblock ``Femtocaching: Wireless video content delivery through distributed
  caching helpers,''
\newblock {\em CoRR}, vol. abs/1109.4179, 2011.

\bibitem{ji2013optimal}
M.~Ji, G.~Caire, and A.F. Molisch,
\newblock ``Optimal throughput-outage trade-off in wireless one-hop caching
  networks,''
\newblock {\em arXiv preprint arXiv:1302.2168}, 2013.

\bibitem{gitzenis2012asymptotic}
S.~Gitzenis, GS~Paschos, and L.~Tassiulas,
\newblock ``Asymptotic laws for joint content replication and delivery in
  wireless networks,''
\newblock {\em Arxiv preprint arXiv:1201.3095}, 2012.

\bibitem{maddah2012fundamental}
M.A. Maddah-Ali and U.~Niesen,
\newblock ``Fundamental limits of caching,''
\newblock {\em arXiv preprint arXiv:1209.5807}, 2012.

\bibitem{sesia-LTE}
S.~Sesia, I.~Toufik, and M.~Baker,
\newblock {\em {LTE}: the Long Term Evolution-From theory to practice},
\newblock Wiley, 2009.

\bibitem{azar201328}
Y.~Azar, G.~N. Wong, K.~Wang, R.~Mayzus, Jocelyn~K. S., H.~Zhao, F.~Gutierrez,
  D.~Hwang, and T.~S. Rappaport,
\newblock ``28 ghz propagation measurements for outdoor cellular communications
  using steerable beam antennas in new york city,''
\newblock in {\em 2013 IEEE International Conference on Communications (2013
  ICC)}, 2013.

\bibitem{daniels201060}
R.~C. Daniels, J.~N. Murdock, T.~S. Rappaport, and R.~W. Heath,
\newblock ``60 ghz wireless: Up close and personal,''
\newblock {\em Microwave Magazine, IEEE}, vol. 11, no. 7, pp. 44--50, 2010.

\bibitem{gupta2000capacity}
P.~Gupta and P.R. Kumar,
\newblock ``The capacity of wireless networks,''
\newblock {\em Information Theory, IEEE Transactions on}, vol. 46, no. 2, pp.
  388--404, 2000.

\bibitem{xue2006scaling}
F.~Xue and PR~Kumar,
\newblock {\em Scaling laws for ad hoc wireless networks: an information
  theoretic approach},
\newblock Now Pub, 2006.

\bibitem{kulkarni2004deterministic}
S.R. Kulkarni and P.~Viswanath,
\newblock ``A deterministic approach to throughput scaling in wireless
  networks,''
\newblock {\em Information Theory, IEEE Transactions on}, vol. 50, no. 6, pp.
  1041--1049, 2004.

\bibitem{franceschetti2007closing}
M.~Franceschetti, O.~Dousse, D.N.C. Tse, and P.~Thiran,
\newblock ``Closing the gap in the capacity of wireless networks via
  percolation theory,''
\newblock {\em Information Theory, IEEE Transactions on}, vol. 53, no. 3, pp.
  1009--1018, 2007.

\bibitem{ozgur2007hierarchical}
A.~Ozgur, O.~L{\'e}v{\^e}que, and D.~NC. Tse,
\newblock ``Hierarchical cooperation achieves optimal capacity scaling in ad
  hoc networks,''
\newblock {\em Information Theory, IEEE Transactions on}, vol. 53, no. 10, pp.
  3549--3572, 2007.

\bibitem{franceschetti2009capacity}
M.~Franceschetti, M.~D. Migliore, and P.~Minero,
\newblock ``The capacity of wireless networks: Information-theoretic and
  physical limits,''
\newblock {\em Information Theory, IEEE Transactions on}, vol. 55, no. 8, pp.
  3413--3424, 2009.

\bibitem{breslau1999web}
L.~Breslau, P.~Cao, L.~Fan, G.~Phillips, and S.~Shenker,
\newblock ``Web caching and zipf-like distributions: Evidence and
  implications,''
\newblock in {\em INFOCOM'99.} IEEE, 1999, vol.~1, pp. 126--134.

\bibitem{wu2010flashlinq}
X.~Wu, S.~Tavildar, S.~Shakkottai, T.~Richardson, J.~Li, R.~Laroia, and
  A.~Jovicic,
\newblock ``Flashlinq: A synchronous distributed scheduler for peer-to-peer ad
  hoc networks,''
\newblock in {\em 2010 Allerton Conference}. IEEE, 2010, pp. 514--521.

\bibitem{molisch2011wireless}
A.F. Molisch,
\newblock {\em Wireless communications},
\newblock 2nd edition, IEEE Press - Wiley, 2011.

\bibitem{li2003linear}
S-Y. Li, R.~W. Yeung, and N.~Cai,
\newblock ``Linear network coding,''
\newblock {\em Information Theory, IEEE Transactions on}, vol. 49, no. 2, pp.
  371--381, 2003.

\bibitem{ho2006random}
T.~Ho, M.~M{\'e}dard, R.~Koetter, D.~R. Karger, M.~Effros, J.~Shi, and
  B.~Leong,
\newblock ``A random linear network coding approach to multicast,''
\newblock {\em Information Theory, IEEE Transactions on}, vol. 52, no. 10, pp.
  4413--4430, 2006.

\bibitem{maddah2013decentralized}
M.~A. Maddah-Ali and U.~Niesen,
\newblock ``Decentralized caching attains order-optimal memory-rate tradeoff,''
\newblock {\em arXiv preprint arXiv:1301.5848}, 2013.

\bibitem{sen1999proxy}
S.~Sen, J.~Rexford, and D.~Towsley,
\newblock ``Proxy prefix caching for multimedia streams,''
\newblock in {\em INFOCOM'99. Eighteenth Annual Joint Conference of the IEEE
  Computer and Communications Societies. Proceedings. IEEE}. IEEE, 1999,
  vol.~3, pp. 1310--1319.

\bibitem{correia2001wireless}
L.~M. Correia,
\newblock {\em Wireless flexible personalized communications},
\newblock John Wiley \& Sons, Inc., 2001.

\bibitem{winner2007d1}
II~WINNER,
\newblock ``D1. 1.2, winner ii channel models,,'' 2007.

\bibitem{karedal2008measurement}
J.~Karedal, A.~J. Johansson, F.~Tufvesson, and A.~F. Molisch,
\newblock ``A measurement-based fading model for wireless personal area
  networks,''
\newblock {\em Wireless Communications, IEEE Transactions on}, vol. 7, no. 11,
  pp. 4575--4585, 2008.

\bibitem{rappaport2011broadband}
T.~Rappaport, F.~Gutierrez, E.~Ben-Dor, J.~Murdock, Y.~Qiao, and J.~Tamir,
\newblock ``Broadband millimeter wave propagation measurements and models using
  adaptive beam antennas for outdoor urban cellular communications,''
\newblock 2011.

\bibitem{joonICC2013}
J~Kim, Y~Tian, S~Mangold, and A.~F. Molisch,
\newblock ``Quality-aware coding and relaying for 60 ghz real-time wireless
  video broadcasting,''
\newblock {\em arXiv preprint arXiv:1304.5315}, 2013.

\end{thebibliography}

\end{document}